\documentclass[preprint,10pt]{elsarticle}
\usepackage{amsmath}
\usepackage{polynom}
\usepackage{amssymb}

\usepackage{color}

\let\mpar=\marginpar
\renewcommand\marginpar[1]{\mpar{\raggedright \scriptsize #1}}
\usepackage[normalem]{ulem}

\usepackage[labelfont=bf,labelsep=period,justification=raggedright]{caption}

\date{}

\usepackage{graphicx}
\usepackage{dcolumn}%
\usepackage{bm}%
\usepackage{hyperref}

\def\be{\begin{equation}}   \def\ee{\end{equation}}

\usepackage{xcolor}

\usepackage{placeins}

\usepackage{xr}

\begin{document}

\begin{flushleft}
{\Large
\textbf{Biological Processes as Exploratory Dynamics}
}
\\
\bf{Jane Kondev$^1$, Marc Kirschner$^2$, Hernan G. Garcia$^{3,4}$, Gabriel L. Salmon$^{5}$, Rob Phillips}$^{5,6}$
\\
(1) Department of Physics, Brandeis University, Waltham, MA, U.S.A.\\
 (2) Department of Systems Biology,
Harvard University, Boston, MA, U.S.A.\\
(3)  Department of Molecular \& Cellular Biology and Department of Physics, Berkeley, California, U.S.A.\\
(4) Chan Zuckerberg Biohub--San Francisco, San Francisco, California, U.S.A.\\
(5)  Division of Biology and Biological Engineering and (6) Department of
Physics, California Institute of Technology, Pasadena, California, U.S.A.\\
E-mail: kondev@brandeis.edu,marc@hms.harvard.edu,hggarcia@berkeley.edu, gsalmon@caltech.edu,phillips@pboc.caltech.edu
\end{flushleft}

\section*{Abstract}
Many biological processes can be thought of as the result of an underlying dynamics in which
the system repeatedly undergoes distinct and abortive trajectories with  the dynamical process only ending when
some specific process, purpose, structure or function is achieved.  A classic example is the way in which microtubules
attach to kinetochores as a prerequisite for chromosome segregation and cell division.  In this example, the dynamics is characterized by apparently futile time histories in which microtubules
repeatedly grow and shrink without chromosomal attachment.   We hypothesize that
for biological processes for which it is not the initial conditions that matter, but
rather the final state, this kind of  exploratory dynamics is biology's unique and necessary
solution to achieving these functions with high fidelity.  This kind of cause and effect relationship can be contrasted to examples from physics and chemistry where the initial conditions determine the outcome.  In this paper, we examine the similarities of many biological processes  that depend upon random trajectories starting from the initial state and the selection of subsets of these trajectories to achieve some desired functional final state.    We begin by reviewing the long history of the  principles of dynamics, first in the context of physics, and then in the context of the study of life. These ideas are then stacked up against the broad categories of biological phenomenology that exhibit exploratory dynamics.  We then build on earlier work by making a quantitative examination of
a succession of increasingly sophisticated models for exploratory dynamics, all of
which share the common feature of being a series of repeated trials that ultimately
end in a ``winning'' trajectory.   We also explore the ways in which microscopic parameters 
can be tuned to alter exploratory dynamics as well as the energetic burden
of performing such processes.\\

It is a great privilege to take part in this special volume dedicated to the life and work of
Prof. Erich Sackmann (1934-2024).  For one of us (RP),  at the time of making a switch from 
traditional condensed matter physics to a life engaged in the study of life, he went to a meeting
near Munich which completely opened his eyes to the ways in which the approach of physics
could be brought to bear on the study of the living.  Sackmann's work was an inspiring presence at that meeting.
One of the hallmarks of his work was a principled approach to dissecting biological processes 
over a range of scales and phenomena.  One common thread to much of his work was that it acknowledged
the dynamical character of living organisms.
The present paper attempts to follow in the tradition of Sackmann's studies of dynamics by suggesting a new way of looking at many biological processes all through
the unifying perspective of what we will call exploratory dynamics.

\section{Theories of Inanimate and Animate Dynamics}

\begin{figure}
\centering{\includegraphics[width=5.3truein]{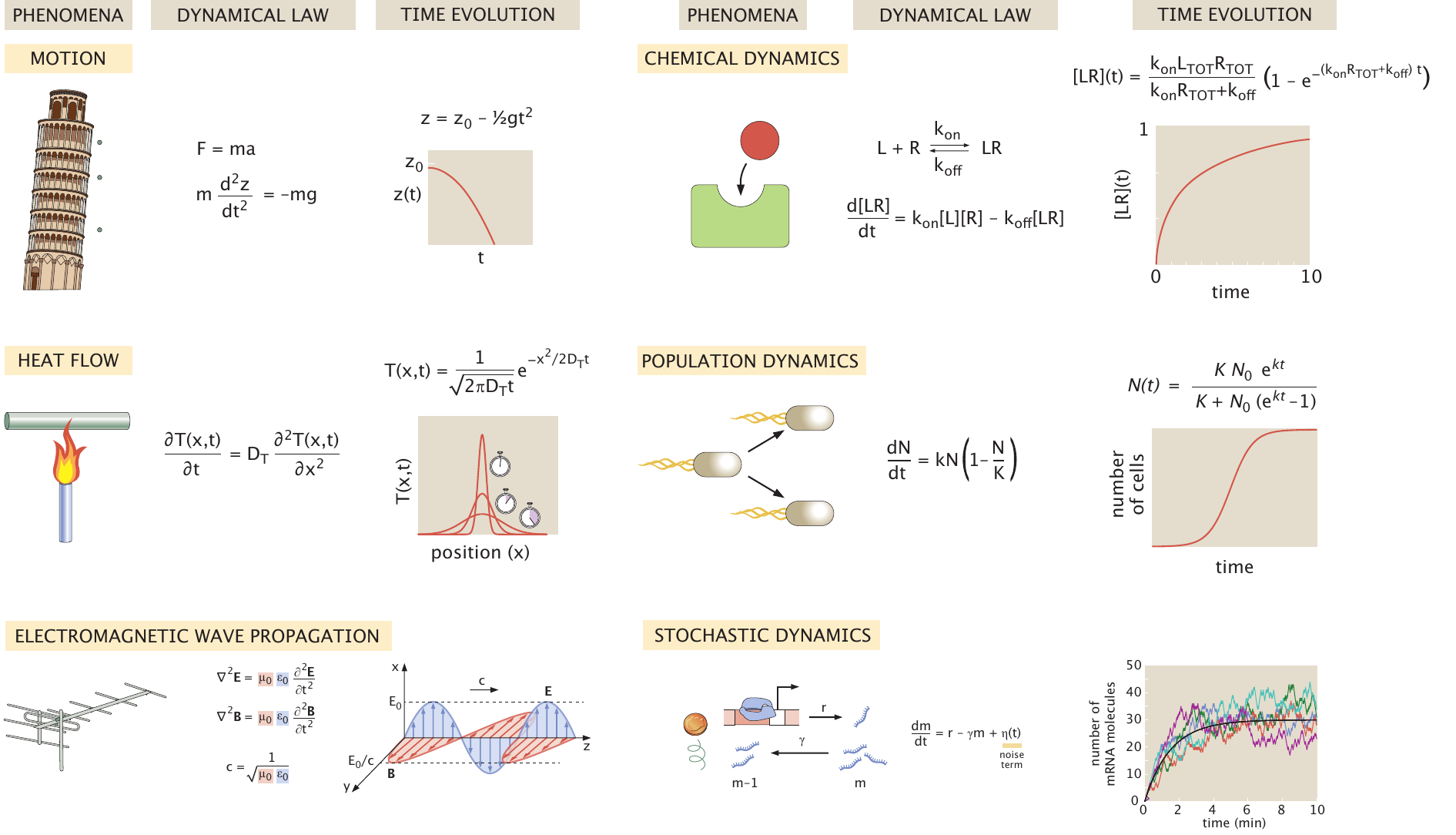}}
\caption{Gallery of dynamical equations.  The defining example of dynamics in physics is the $F=ma$ dynamics of Newton that allows us to solve problems such as falling bodies or planetary motion. The dynamics of chemical reactions is another well established example of initial-condition driven dynamics, in this case
showing the dynamics of ligand (red)-receptor (green) binding.
 The heat equation is a 
deterministic dynamical equation that tells us how an initial temperature distribution will evolve in space and time. 
There are a host of different dynamical equations describing the dynamics of populations, here we show the logistic equation.
Maxwell's theory of is a quintessential example of a field theory which vastly expands the repertoire of classical physics.   Langevin-like equations even allow us to work out the time evolution of
stochastic chemical dynamics.
\label{fig:GalleryInitialConditionDynamics}}
\end{figure}

Whatever level of phenomena we look at, small or large spatial scales, short or long time scales, low or high energies, these phenomena are usually dynamic.  
The quest to understand
the dynamics of broader and broader classes of phenomena has repeatedly over the long arc of
the history of science
resulted in ``new physics''~\cite{Einstein1961,Brush1986,Segre2007,Longair2020}.  
Indeed, as shown in Figure~\ref{fig:GalleryInitialConditionDynamics}, there is a long and justly celebrated history of different approaches
to working out the dynamics of processes in the world around us.   
The template for this subject both mathematically and physically was achieved by Newton standing on the shoulders of Galileo and
others~\cite{Cohen1981}.    Newton saw that ${\bf F}=m{\bf a}$ provided an update rule that allowed
us, using what we now know as calculus, to find the current state of
the system by time stepping from previous states~\cite{Pask2019}.  Though most of us learn about
Newton's analysis of the planets in polar coordinates culminating in elliptical orbits, Newton's treatment
of the problem was as a series of straight line steps punctuated by impulses due to the force of gravitation resulting in an orbit that was a polygon~\cite{Pask2019}.  We might summarize this and most subsequent dynamical
laws as having the form
\begin{equation}
\mbox{state}(t+\Delta t)=\mbox{state}(t) + \mbox{update}(t) \Delta t,
\end{equation}
the simplest embodiment of the forward-Euler method for solving differential equations,
but which we think of differently as being the discrete representation of most versions
of deterministic, initial-condition dictated,  dynamics.  The state now is determined by the
state a time $\Delta t$ earlier plus some update term.

 In the centuries that followed Newton's creation of the ``System of the World~\cite{Pask2019},'' $F=ma$ dynamics was brilliantly generalized to continuous media with examples such as
hydrodynamics and elastodynamics to describe the mechanics of fluids and solids~\cite{Thorne2021,HueschenBook2024}.  But other phenomena called for 
something beyond the mechanical world view~\cite{Klein1973}.
As shown in Figure~\ref{fig:GalleryInitialConditionDynamics},
 new dynamical laws were articulated that made it
possible to work out the time evolution of a myriad of processes.  
For example, the study of chemical transformations led to the emergence of 
 the law of mass action and the rate equations of chemical kinetics,
allowing for the appearance and disappearance of different chemical species~\cite{Kudryavtsev2001,Feinberg2019}.  Another class of problems highlighted in
Figure~\ref{fig:GalleryInitialConditionDynamics} arose as Fourier and others
tackled the  dynamics of temperature fields and heat flow in the form of
a partial differential equation (the heat equation) showcased in the figure~\cite{Fourier2003, Brush1986}.  In this case, a material
subjected to some initial temperature distribution will evolve to some new temperature distribution~\cite{Crank1975,Carslaw1986}. 
The dynamical equations describing the diffusion of matter and heat had and have a huge reach including
to important problems such as the age of the Earth~\cite{Richter1986,Dalrymple1991,Dalrymple2004}.
While physicists were busy working out the laws of inanimate matter, other kinds of scientists and mathematicians
were also engaged in thinking about the dynamics of populations as exemplified by the logistic equation shown  in the figure~\cite{Lotka1956,Case2000,Provine2001,Bacaer2011}.
In the middle of the 19th century, laws of dynamical evolution received a great boost in the form of
Maxwell's theory of the electromagnetic field to describe  the spacetime evolution of electromagnetic fields~\cite{Einstein1961,Segre2007,Longair2020}. In addition to these exercises in deterministic dynamics, the discovery of Brownian motion inched scientists 
and mathematicians towards the consideration
of random processes such as the jiggling motions of pollen grains observed by Brown, in that case using the Langevin equation for example~\cite{Chandrasekhar1943,Lemons2002}.
Many of these dynamical descriptions share the common feature that 
there is some initial state of the system, and the dynamical laws tell us
how to compute the subsequent evolution of the system in time. 

But the dynamical laws described above left out one of the most important
categories of dynamics in the natural world, namely, the strange dynamics in and of living organisms, processes including the long chains of energy-driven biochemical reactions, structure formation from macromolecular complexes to organelles, embryonic development and a myriad of other fascinating topics.  It is fashionable to minimize the brilliant efforts of our predecessors when we later come to realize 
how ``wrong'' their ideas might have been.  Within physics the 19th century idea of heat as a fluid,
the famed caloric, or of the luminiferous ether, the medium within which electromagnetic disturbances were thought to propagate, are sometimes viewed as quaint and misguided. However, a more fair and nuanced view sees them instead as brilliant and creative examples of the continuing refinement that attends most scientific progress~\cite{Einstein1961,Segre2007,Longair2020}.  In the study of evolution, Lamarck's ideas
are routinely dismissed as similarly misguided, though a more scholarly judgment is that like many others, Lamarck was engaged in thoughtful hypothesis making that was later learned to be inconsistent with
observations and measurements.
Perhaps no discredited topic draws such contempt as does that of vitalism.  However, even here, a more
generous interpretation of the 19th century vitalists is to acknowledge the obvious: living organisms
seem different than their inanimate counterparts and clearly, the incomplete dynamical repertoire of classical physics might have something to gain by coming to terms with the purposeful dynamics of living organisms.

\begin{figure}
\centering{\includegraphics[width=5.3truein]{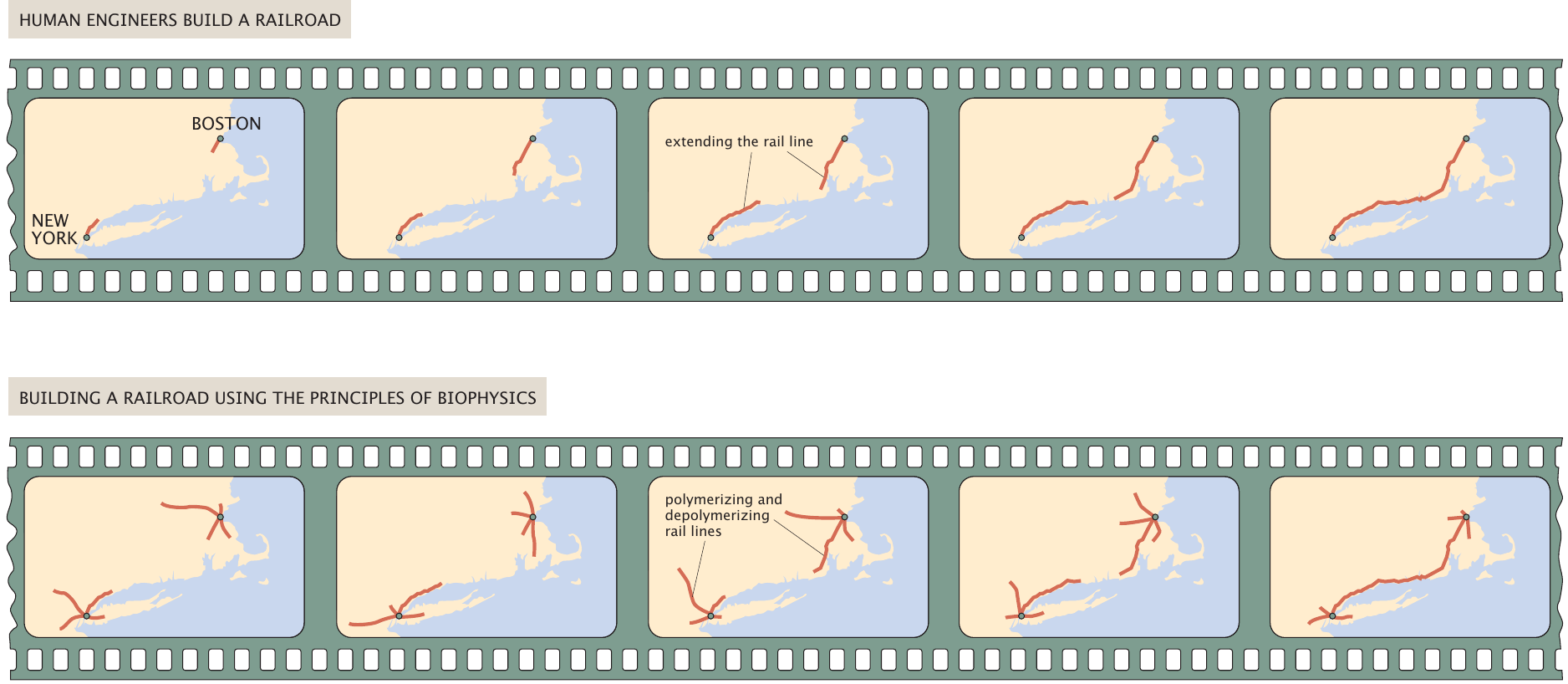}}
\caption{Comparison of human-scale dynamics and cellular-scale dynamics approaches to building
a railroad between New York City and Boston. In both cases, there is a very specific goal.  However, the strategy for achieving that goal is completely different in the human engineering context in comparison with the 
biological strategy of using stochastic variation and ``selection''  to achieve exploratory dynamics.  The concept of  the figure is
due to a suggestion of Andrew Murray.
\label{fig:NYBostonRail}}
\end{figure}

\begin{figure}
\centering{\includegraphics[width=5.3truein]{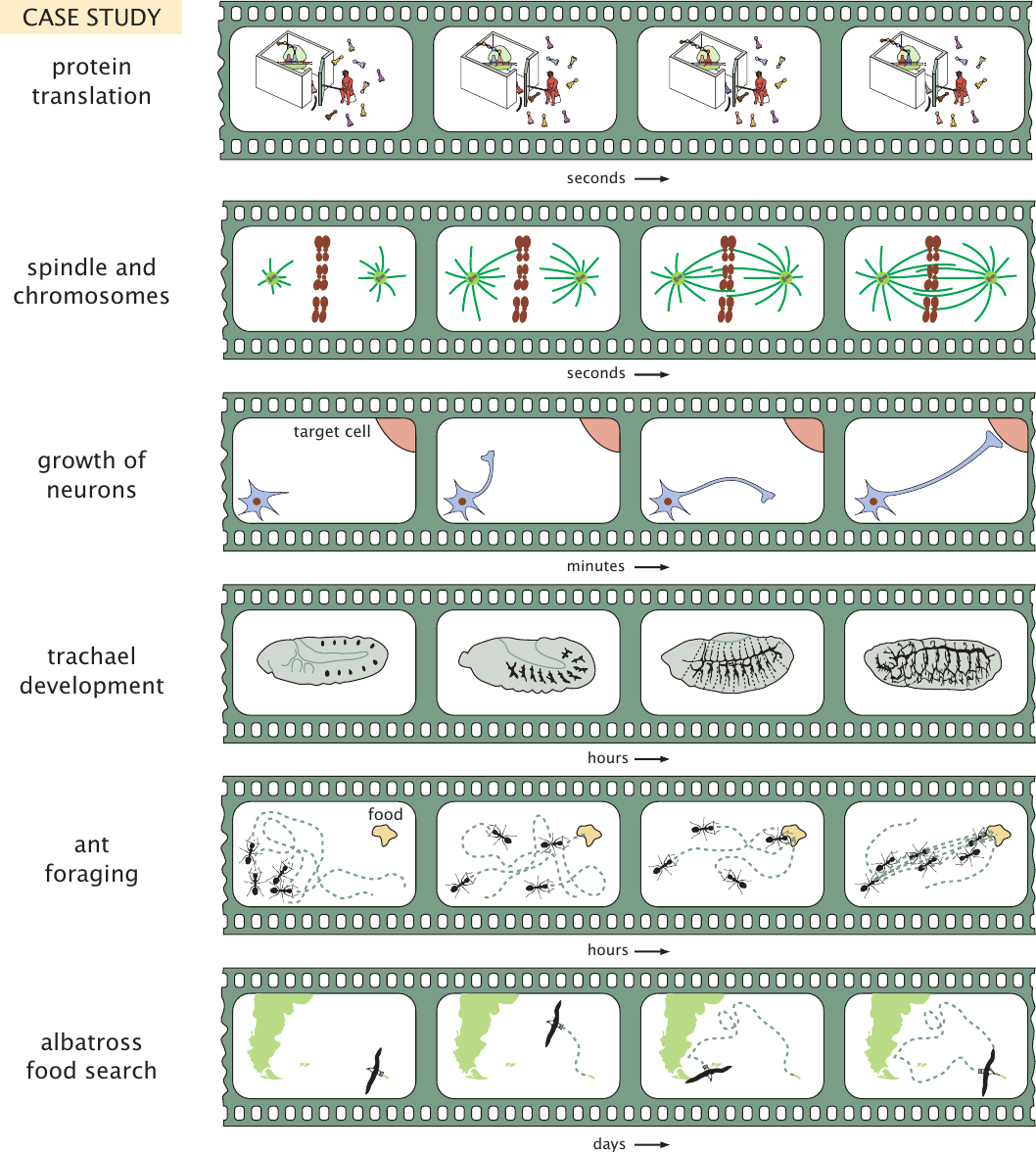}}
\caption{Gallery of exploratory dynamics.  Examples from a wide variety of spatial and
temporal scales illustrate the way in which the dynamics is characterized by a set of trajectories,
all of which repeatedly fail to achieve their function before ultimately succeeding.  Each example
is explained in more detail in the text.
\label{fig:GalleryExploratory}}
\end{figure}

One way in which the study of the living might hold the secret to new principles is
in the sense that the paradigm of Laplace's Demon~\cite{Canales2020}, the strict unfolding of
initial conditions through deterministic evolution, is often not respected  by processes within living cells.  Au contraire, 
cells often require a different kind of dynamics that is driven by
the {\it final} conditions rather than the {\it  initial} conditions and it is to this kind of
processes that the remainder of our paper is dedicated.     In particular, living organisms have developed
ways of converting goals or functions into goal-oriented dynamical processes.  In many everyday human processes characterized by
some purpose, goal or function, the standard paradigm is the construction of machines.   We build machines
to pump water, to cool materials~\cite{Twilley2024}, to move objects and the list goes on and on.  By way of
contrast, living
organisms have used stochasticity in a novel way to project the future into the past,  to enable random 
trajectories to ``find'' the goal or target  in a process which 
we call exploratory dynamics~\cite{Gerhart1997}. 
Andrew Murray suggested to
us an everyday occurrence that might help frame the uniqueness and weirdness of
exploratory dynamics as shown in Figure~\ref{fig:NYBostonRail}.  He asks us to imagine building a railroad line between
New York City and Boston.   We all know how this goes if done according to
the strictures of modern civil engineering. There is a goal, there is a plan, the plan is implemented
to achieve the goal.  Surveyors do their work, steel is ordered, worked are deployed and paid to lay down track.  However, if done according
to the principles of biological exploratory dynamics, instead, construction workers will
start in the two cities, randomly choosing a direction to start laying down track. If after a certain
time, they have not encountered their counterparts coming from the other city, they
remove the track and try again.  That's all.    For most of us, our immediate reaction is
that this is woefully inefficient, perhaps even absurd.   However, as seen in the examples of Figure~\ref{fig:GalleryExploratory},
this kind of dynamics describes important biological processes over a dizzying array of spatial and temporal scales.    Though this idea of exploratory dynamics was first clearly articulated
in the context of the search and capture mechanism of chromosome segregation~\cite{KirschnerSearch1986}, in the time since a growing list of examples of
this kind of dynamics has mounted~\cite{Gerhart1997}.  As a result,  we think we have reached a moment
of timely rather than premature  abstraction in which experience with these specific examples suggests that instead of
being a particular mechanistic feature of chromosome search, exploratory dynamics
is a much more general phenomenon, calling for its own abstractions, toy models and equations.
 In this paper, we examine one category of exploratory dynamics
which features a series of abortive trajectories followed by a final successful trajectory.

The organization of
the paper is as follows.  In section~\ref{section:ExploratoryDynamics} we explain the
concept of exploratory dynamics, first by describing biological processes that exemplify it
and then by giving a cursory description of the kinds of theoretical ideas that
have been put forth that might help explain such processes.  Section~\ref{section:ZeroD} explains the statistical
mechanics of exploratory trajectories using a simple zero-dimensional model. With
those preliminaries in hand, section~\ref{section:ChromosomeSearch} describes
the concrete problem of microtubules searching for chromosomes in the language
of exploratory dynamics.  Section~\ref{section:TFSearch} tackles a completely different
biological phenomenom, namely, how transcription factors find their target site on DNA,
revealing many mathematical similarities to
the problem of the microtubule search problem.  We note that in both of our case studies, we in no way pretend that we are the first to write down many of the models we explore, nor do we pretend our limited survey of the relevant literature to be comprehensive or encyclopedic.  Rather, we spell them out in full detail in the hope of carefully illustrating the statistical mechanics of exploratory trajectories in a few limited examples in the hope that it might encourage 
analysis of the full suite of examples shown in Figure~\ref{fig:GalleryExploratory}, with the aim of more mathematical generality than the few examples considered here.  We finish in section~\ref{section:Discussion} with a high-level summary of the possible broader implications of exploratory dynamics and some final reflections
on the career and influence of Prof. Erich Sackmann.

\section{Exploratory Dynamics by Variation and Selection}
\label{section:ExploratoryDynamics}

In many cases, when we use or hear the words variation and selection, it
is in the context of Darwinian evolution.  However, the dynamics of
variation and selection has much broader reach than in the context of
traditional biological evolution. Perhaps surprisingly, there are many problems in molecular and cellular dynamics that can be couched in this same language.    Examples range from the search of transcription factors for
their binding sites~\cite{vonHippel1979,Winter1981a,Winter1981b,Hu2006,Hachmo2023}; microtubules being captured by kinetochores on chromosomes~\cite{Mitchison1985,Holy1994,Gundersen2002,Wollman2005,Heald2015};
bacterial cells performing the biased random walks of chemotaxis~\cite{Berg1993,BergBook2004}; the formation of
vasculature in developing embryos~\cite{Ghabrial2003}; the formation of exquisitely precise neuronal connections~\cite{Li2021}; the development of antibody repertoires~\cite{Tonegawa1983,Tauber1997,Janeway2012};
  animals forging for food sources~\cite{Prince1992,Block2011,Viswanathan2011,Kays2015,Antolos2017,Hindell2020} and far beyond.  In an earlier work, one of us (MK) in collaboration with John Gerhart wrote a book which considered living organisms in terms of some key
  principles that might describe broad swathes of biological phenomena~\cite{Gerhart1997}.  Chapter 4 of that book entitled ``The Exploratory Behavior of Biological Systems'' described a series of case studies in biological phenomena over a wide range of spatial and temporal scales that could be viewed as being of an exploratory character.  Those examples inspired the search for 
 biophysical underpinnings to exploratory dynamics, and this paper is is our attempt to lay out
 several case studies in some mathematical detail.
  
Figure~\ref{fig:GalleryExploratory} provides
a gallery of examples of biological processes that we think it might be fruitful to examine quantitatively from the perspective of exploratory dynamics.  
Though the formal definition of what constitutes exploratory dynamics might still be coming into focus, for the purpose of the present paper, we consider a process as exploratory if it is characterized by a well defined final state which is reached through a series of trials.  In particular, we imagine a random process that produces a set of trajectories, a ``standing variation'' in trajectory space.  A selection mechanism then  picks out a subset of these trajectories, which are ``functional.''  One can imagine refinements in which we ask whether the exploratory process
requires energy investment such as in the case of chromosome search and capture, or not, as in the case of  transcription factor search as considered later in the paper.  In addition, one can also imagine situations in which there are cues that reinforce some trajectories that appear to be headed in the ``right'' direction such as is found in chemotaxis or the growth of vasculature.

\begin{figure}
\centering{\includegraphics[width=6.2truein]{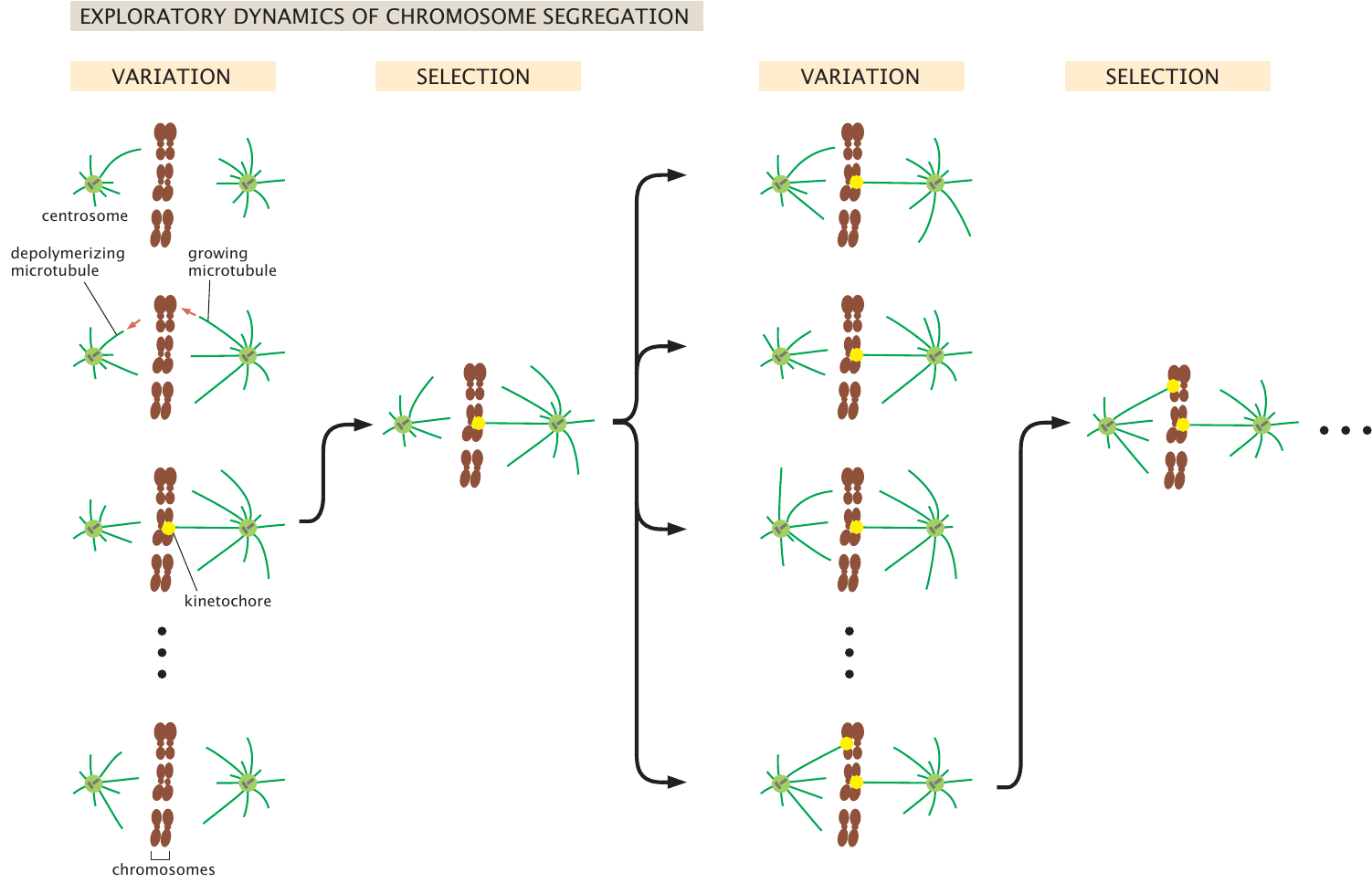}}
\caption{Microtubule search for kinetochores as exploratory dynamics characterized
by stochastic variation in the dynamical trajectories of the system coupled with selection and stabilization
of those trajectories that culminate in attachment to kinetochores.
\label{fig:KinetochoreVariationSelection}}
\end{figure}

To more concretely illustrate what we mean by exploratory dynamics, the first panel of Figure~\ref{fig:GalleryExploratory} reminds us that the process of protein translation
is exploratory in the sense that repeatedly, the ``wrong'' amino acid is presented to the ribosome
and only through this kind of biased trial-and-error dynamics does a protein get translated
with high fidelity~\cite{Hopfield1974,Ninio1975}.  In most cases in which a wrong tRNA arrives at the ribosome, there is a reset that restores the system back to its starting state.  In this case, the ``variation'' refers to at least at 20:1 ratio of wrong to right tRNAs arriving at the ribosome, and the ``selection'' refers to the incorporation of the amino acid of interest into
the nascent polypeptide chain.
As shown in the second panel, one of the most compelling examples of
exploratory dynamics  is the dynamics of microtubules which precedes chromosome segregation in dividing
cells~\cite{Mitchison1985,Holy1994,Gundersen2002,Wollman2005,Heald2015}.  Here, microtubules grow and shrink repeatedly from an organizing center known
as the centrosome, and those microtubules that find the kinetochore on the chromosomes
stay attached to their target.   A reframing of the problem as one in exploratory dynamics is
schematized in Figure~\ref{fig:KinetochoreVariationSelection}.
The third example shows the growth of neurons, where  connections come and go
and are ``selected'' in some contexts as illustrated in the example of olfactory
selection cited here~\cite{Li2021}.  In the developing fly embryo, though these animals
don't have vasculature in the same sense as mammals, they do have a system of ``pipes'' that allow
for the transport of oxygen to cells and the establishment of this spatial arrangement of
pipes is itself exploratory~\cite{Ghabrial2003}.   More generally, the subject of branching morphogenesis 
provides a broad array of examples of exploratory dynamics~\cite{Hannezo2019,Palavalli2021}.   At much larger scales; animal foraging can be thought of
as its own kind of exploratory dynamics~\cite{Prince1992,Block2011,Viswanathan2011,Kays2015,Antolos2017,Hindell2020}.   The main point of the figure and this discussion is
simply to capture the reader's imagination with the hypothesis that perhaps exploratory
dynamics is a new and important part of the dynamics repertoire that has thus far been
understudied in general terms from the physics perspective.  
  
Our discussion thus far primarily focused on the biological phenomenology of exploratory
dynamics.  Interestingly, in parallel, the worlds of both physics and mathematics had
developments of their own aimed at solving problems in
purposeful dynamics.  Though it might seem far afield, the quest of the Allied Forces to search for German
U-Boats gave rise to the inspiring and useful field of operations research and search theory~\cite{Koopman1980,Wagner1999,Jaynes1981}.
Ironically, several of the 20th century's most famous biologists were key participants
in the development of the mathematical and physical tools to treat search problems~\cite{Budiansky2013}, including
John Kendrew, noted for his resolution of the molecular structure of whale myoglobin~\cite{Judson1996}.
Perhaps more surprisingly is that at the age of 67, CH Waddington, of
Waddington landscape (and many other things) fame, wrote a profound text on
the use of operations research during the Second World War of which he, along with Kendrew, was a key participant~\cite{Waddington1973}.  A parallel avenue of theoretical research that we think is
quite relevant to exploratory dynamics has focused on random walks with resets~\cite{Bressloff2020,Evans2020}.  Yet another set of ideas were introduced that have been dubbed 
infotaxis~\cite{Vergassola2007,Loisy2022,Monthiller2022}.  In the remainder of the paper, we focus on
a specific approach in which we describe exploratory dynamics as a statistical mechanics
of a particular class of trajectories, but we are anxious to see what theorists will have to say about
exploratory dynamics in coming years.

\section{Zero-Dimensional Exploratory Dynamics}
\label{section:ZeroD}

As noted above, one example of exploratory dynamics that is enormously consequential is the way in which
microtubules pair up with chromosomes in preparation for the process of
chromosome segregation as shown schematically
in Figure~\ref{fig:KinetochoreVariationSelection}.  In this paper, we use this example as a specific case study
that illustrates what are perhaps more general ideas of how to formulate exploratory
dynamics as a statistical mechanics of trajectories.  
In this  statistical mechanics of exploratory trajectories, all trajectories are characterized by
the fact that they end when some ``purpose'' is fulfilled or some function is achieved. 
As we show below, in each case considered in this paper, this results in the use of the geometric
distribution.  We suspect that there are much more general ideas waiting to be discovered
that can account for cases in which subsequent trajectories have a memory of those that have come before or in which signals reinforce trajectories that are headed in the right direction.

\begin{figure}
\centering{\includegraphics[width=4.3truein]{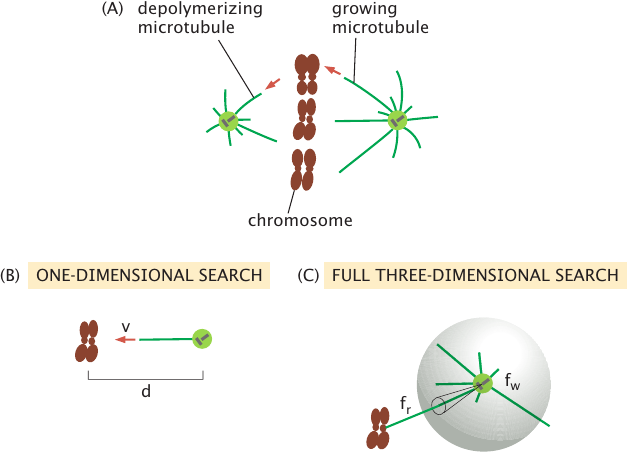}}
\caption{Schematic of the process of microtubules searching for chromosomes. (A) Schematic showing how microtubules grow out from the centrosome and then suffer catastrophes.  (B) One-dimensional idealization of the process of
microtubules searching for a chromosome.  We imagine that the distance from the centrosome to the kinetochore is $d$ and that the polymerizing microtubule grows with speed $v$.  All chromosomes grow in the same direction and the exploratory character is in how long the microtubule survives before suffering a catastrophe.  (C) Three-dimensional idealization of the process of
microtubules searching for a chromosome.  $f_r$ is the probability of the microtubule growing in the right direction and $f_w = 1-f_r$ is the probability of growing in the wrong direction.  The right direction is defined by a cone of solid angle such that the microtubule will grow in a range of allowed directions that will contact the kinetochore.
\label{fig:ChromosomeSchematic}}
\end{figure}

As a first model, we set up the mathematical language of
the statistical mechanics of exploratory trajectories by considering 
trajectories built up of ``wrong'' and ``right'' moves.  For
the moment, we are noncommittal about the moves themselves
since as we will see later in the paper, in different contexts, ``move'' will refer
to the duration of the polymerization process,  the direction of polymerization or to changing
between one-dimensional and three-dimensional diffusion in the case of transcription factors
searching for their binding sites.    All the trajectories
have the same basic structure.  A series of wrong moves followed by a final right
move.  At each time step,
the wrong and right moves are each chosen with probability $f_w$ and
$f_r$, respectively, subject to the condition that $f_w+f_r=1$.  A given exploratory trajectory is assembled as a collection of  $i$ unsuccessful steps before a final successful step.  The probability of such a trajectory is given by 
\begin{equation}
p_i=\underbrace{f_w f_w \cdots f_w}_{\text i} f_r  = f_w^i f_r.
\end{equation}
We can rewrite this more intuitively as 
\begin{equation}
p_i = (1-f_r)^{i} f_r,
\end{equation}
a probability distribution known more colloquially as the geometric distribution~\cite{Blitzstein2019}.
This fundamental equation has a structure that we will use over
and over in thinking about the statistical mechanics of exploratory
trajectories since it highlights the importance of the final state.

Since we will need to manipulate these distributions repeatedly, here we use
it to compute the average waiting time until the exploration is complete.
We begin by verifying that the distribution is normalized by computing
\begin{equation}
\sum_{i=0}^{\infty} p_i= \sum_{i=0}^{\infty}(1-f_r)^{i} f_r =f_r   \sum_{i=0}^{\infty}(1-f_r)^i.
\end{equation}
  We recognize the  remaining sum as a geometric
series with the property that  $\sum_{i=0}^{\infty} x^i=1/(1-x)$ if $0 < x < 1$.   Since we have
$0 < 1-f_r < 1$,  our case yields
\begin{equation}
 \sum_{i=0}^{\infty} p_i=f_r \sum_{i=0}^{\infty}(1-f_r)^i ={f_r \over (1-(1-f_r))}=1.
\end{equation}
As expected, the distribution is normalized. 

A more important question is how long does it take
on average for the exploratory dynamics to terminate due to success in finding its
target?
The answer to that question is obtained by averaging over the time spent in all
the possible trajectories and given by
\begin{equation}
\langle t\rangle=\sum_{i=0}^{\infty} \underbrace{\left(i \, t_{\text {step }}\right.}_{\text {failures }}+\underbrace{t_{\text {step}}}_{\text {success}}) p_i,
\label{eq:firstAvgExploreTime}
\end{equation}
where we have introduced the quantity $t_{\text{step}}$ as the duration of each instance where the system makes an unsuccessful excursion (e.g. a microtubule polymerizes without finding the chromosome).
Later in
the paper, we will explicitly calculate $t_{\text{step}}$ on the basis of the microscopic
parameters associated with the microtubule polymerization problem or the
transcription factor search problem.
The second term in the sum immediately yields $t_{\text{step}}$ since the distribution is normalized
leaving us with
\begin{equation}
\langle t\rangle=\sum_{i=0}^{\infty}i \, t_{\text{step}}\, p_i+t_{\text{step}}.
\end{equation}
Next, we need to evaluate
\begin{equation}
\langle t \rangle = t_{\text{step}}\sum_{i=0}^{\infty} i \, p_i+t_{\text{step}}=  t_{\text{step}} \, f_r \sum_{i=0}^{\infty} i  (1-f_r)^{i}+t_{\text{step}},
\label{eqn:AverageTime1}
\end{equation}
which we recognize as having a term of the form
\begin{equation}
\sum_{i=0}^{\infty} i x^i = x {d \over dx} \sum_{i=1}^{\infty} x^i  = x {d \over dx}  {1 \over 1-x}= {x \over (1-x)^2}.
\label{eqn:DerivativeTrick}
\end{equation}
We now see that we can rewrite eqn.~\ref{eqn:AverageTime1} as
\begin{equation}
\langle t \rangle = t_{\text{step}} \, f_r \sum_{i=0}^{\infty} i  (1-f_r)^{i}+t_{\text{step}}=t_{\text{step}} \, (1-x) \sum_{i=0}^{\infty} i  x^{i}+t_{\text{step}},
\end{equation}
where we have used the definition $x=1-f_r$. We now invoke the derivative trick introduced in eqn.~\ref{eqn:DerivativeTrick} to arrive at
\begin{equation}
\langle t \rangle = t_{\text{step}} \, (1-x) {x \over (1-x)^2} + t_{\text{step}} = t_{\text{step}} \, {x \over (1-x)} + t_{\text{step}}.
\end{equation}
If we now write this expression in terms of $f_r$, we find
the very intuitive result that 
\begin{equation}
\langle t \rangle =t_{\text{step}}\left(1 + {1-f_r \over f_r}\right)= {t_{\text{step}} \over f_r}.
\end{equation}
The intuition for this result is best served by concretely imagining that $f_r=1/M$ which means
that $1$ out of every $M$ trials is successful.  This implies in turn that $\langle t \rangle=M \, t_{\text{step}}$,
stating that on average we need to carry out $M$ trials for the exploration
to be a success when the probability of success on a given trajectory is $f_r=1/M$.

Our reason for going through this example is to showcase the structure of the result,
 since in all of our
examples of exploratory dynamics, we will ultimately characterize the exploratory
process as a statistical mechanics of trajectories, with all such trajectories 
sharing the feature that they involve repeated ``failures'' until they succeed on
the final trajectory.  Interestingly, for the examples considered throughout
the paper, we will see many different realizations of the space of trajectories,
but as noted above, they all are of the form $p_i=\underbrace{f_w f_w \cdots f_w}_{\text i} f_r$.

\section{Chromosome Search Via Microtubule Reset Dynamics}
\label{section:ChromosomeSearch}

\subsection{One-Dimensional Chromosome Capture by Exploratory Dynamics}

As our next case study, we consider the one-dimensional model
summarized in Figure~\ref{fig:ChromosomeSchematic}(B) where we idealize
the growth and catastrophes as taking place in one dimension and ask for
the time it takes for the microtubule that is subject to both growth and shrinkage
to find its chromosomal target.  This analysis largely imitates the beautiful work of Holy and Leibler,
especially as developed in their Appendix B~\cite{Holy1994} as well as more recent work
from Wollman et al.~\cite{Wollman2005} and their thorough and insightful
Supplementary Information.   In the spirit of successively building up models of
increasing sophistication, we begin with the unrealistic assumption that every microtubule
grows in the right direction and ask only whether it survives long enough to
reach the chromosome.   One way to think about this model is that the exploratory 
behavior only exists in the sense that the different exploratory periods have
different durations.  Later, we will extend this framework to consider the
case in which there is also a spatial component to the search process
with different microtubules heading off in different directions, lending
a second facet to the exploratory nature of the dynamics.

Before we can compute the time scale associated with the
chromosomal search process, we must first consider the waiting time distribution
that tells us how long a microtubule grows before it suffers a catastrophe.  
To answer that question, we turn once again to the geometric distribution. 
If the rate constant for a catastrophe is $k_c$, then the probability in a time
step $\Delta t$ that the microtubule will {\it not} suffer a catastrophe is
$(1-k_c\Delta t)$.    Hence, if the microtubule is to pass through $N$ time steps before having
a catastrophe, the probability of that succession of events is
\begin{equation}
p_N = (1-k_c\Delta t)^N k_c \Delta t.
\end{equation}
Because of the beautiful properties of this distribution in the large $N$ limit, namely
\begin{equation}
\lim_{N\rightarrow \infty} \left( 1- {x \over N} \right)^N = e^{-x},
\end{equation}
and using the fact that the total time elapsed is  $t=N\Delta t$, we see
that the probability that the microtubule will suffer a catastrophe
in the time interval between $t$ and $t+dt$ is given by
\begin{equation}
p_{\text{survive}}(t) dt = k_c e^{-k_c t} dt.
\end{equation}
The most naive model of the average time before a microtubule suffers a 
catastrophe is given by $t_{c}=1/k_c$ which we find by evaluating the average
\begin{equation}
t_c= \int_{0}^{\infty} t k_c e^{-k_c t} d t ={1 \over k_c}.
\end{equation}
However, the average time before a catastrophe is actually shorter.
Note that if the distance from the initial position of the microtubule to the chromosome is $d$ and
it grows at a rate $v$, this implies that if the microtubule grows for a time longer than $\tau=d/v$, the
microtubule will have hit the chromosome, thus ending the exploratory set of trajectories.
  Hence, we need to be more careful in assigning a time
scale to the failed trajectories.  Instead, we should sum only over those trajectories that last a time
shorter than $\tau$.  In particular, the average time of the failed trajectories is given by
\begin{equation}
t_f=\frac{\int_0^{\tau} e^{-k_c t}k_c t d t}{\int_0^{\tau} e^{-k_c t} k_c d t},
\end{equation}
where we only allow those trajectories that are shorter than $\tau$
and we have introduced the notation $t_f$ with subscript $f$ to denote
``failed'' trajectories. The denominator imposes normalization on the distribution.
These integrals can be evaluated to yield
\begin{equation}
t_f=\frac{1}{k_c} \frac{1-\left(k_c \tau+1\right) e^{-k_c \tau}}{1-e^{-k_c \tau}}.
\label{eqn:tr}
\end{equation}
Note that this average time until a catastrophe is shorter than the naive estimate
of $1/k_c$ derived above.

\begin{figure}
\centering{\includegraphics[width=4.4truein]{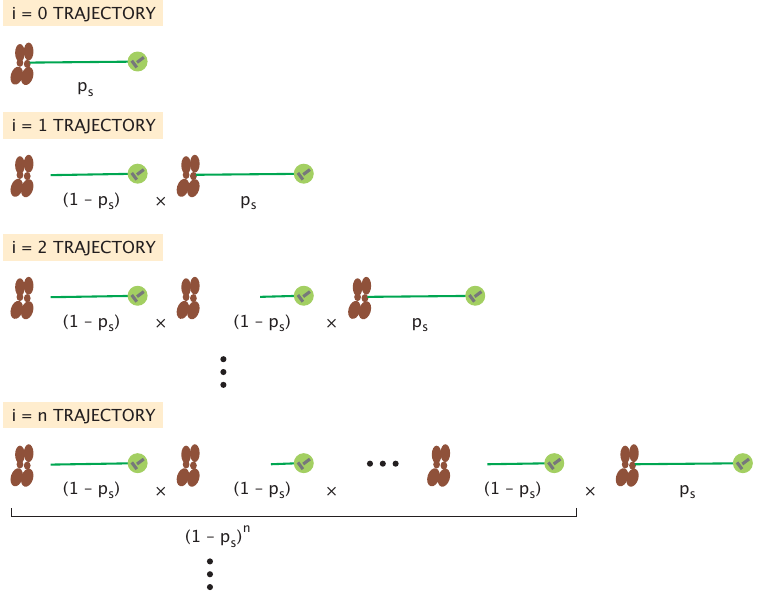}}
\caption{Trajectories and weights for one-dimensional model of chromosome search and capture.
The index ``$i$'' labels the number of times that the microtubule {\it fails} to hit the chromosome before
its eventual success.  The ``failed'' trajectories are characterized by the fact that the microtubule suffers a catastrophe before reaching the chromosome.  The failed trajectories show the maximum length of the microtubule before suffering a catastrophe.
\label{fig:ChromosomeSearch1D}}
\end{figure}

Given the survival  distribution, the probability
that the microtubule will survive long enough to reach the chromosome is given by
\begin{equation}
p_s=\int_{\tau}^{\infty} k_c e^{-k_c t} d t =e^{-k_cd/v}.
\end{equation}
The integral computes the fraction of the probability distribution where the survival of the microtubule lasts
longer than the time to reach the chromosome, $\tau=d/v$, where again, $d$ is the distance
to the microtubule and $v$ is the velocity of microtubule growth. 
Our goal is to find the average time it takes for the growing and shrinking microtubule
to reach the chromosome.  The probability of success after $i$ failures is once again given by
the geometric distribution which we now write as 
\begin{equation}
p_i=\left(1-p_s\right)^{i} p_s
\label{eqn:successithtrial}
\end{equation}
which reflects the fact that for $i$ of the growth processes, the microtubule
suffers a catastrophe before reaching the chromosome and only on the $i+1^{th}$ try does
it reach the chromosome with probability $p_s$. A variety of examples of these kinds of exploratory
trajectories are shown in Figure~\ref{fig:ChromosomeSearch1D}.   The time taken for this particular class of trajectories
is given by $\tau + i t_f$ since on average each of the failed growth and catastrophe events
takes a time $t_f$ and the final successful growth event takes a time $\tau$.

\begin{figure}
\centering{\includegraphics[width=3.3truein]{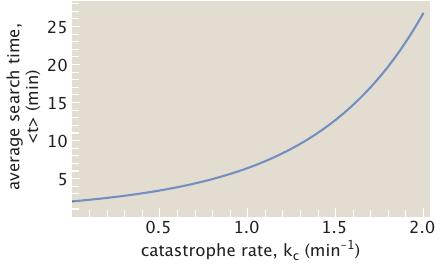}}
\caption{Search time as a function of the catastrophe rate for the one-dimensional model of chromosome search and capture.  For this case $\tau = 2$~min.
\label{fig:ChromosomeSearchTime1D}}
\end{figure}

The logic of our calculation of the average ``exploration time'' is that we need to
sum over all trajectories, every possible life history of the growing and shrinking microtubules
that ends by ultimately finding the chromosome.  In words, this means we have to sum over
the case where the microtubule finds the chromosome on the first try, the second try, the third try
and so on.  
Given this procedure, the average time to find the chromosome is given by
\begin{equation}
\langle t\rangle= \sum_{i=0}^{\infty} (\tau + i t_f)p_i.
\label{eq:secondAvgExploreTime}
\end{equation}
Using the probability of success on the $(i+1)^{th}$ trial introduced in eqn.~\ref{eqn:successithtrial},
we can rewrite the average exploration time as
\begin{equation}
\langle t\rangle= \sum_{i=0}^{\infty} (\tau + i t_f)\left(1-p_s\right)^{i} p_s,
\end{equation}
which is helpfully rewritten as
\begin{equation}
\langle t\rangle= \sum_{i=0}^{\infty} \tau\left(1-p_s\right)^{i} p_s
+\sum_{i=0}^{\infty} i t_f\left(1-p_s\right)^{i} p_s.
\end{equation}
The first of these sums yields $\tau$ since the geometric distribution is
normalized.  For the second sum, 
by exploiting the very useful derivative protocol for
evaluating such sums that we already introduced in eqn.~\ref{eqn:DerivativeTrick}, 
we can write the second sum as
\begin{equation}
\sum_{i=0}^{\infty} i \, t_f\left(1-p_s\right)^{i} p_s =t_f \, p_s  {(1-p_s) \over p_s^2}.
\end{equation}
Hence, the total search time is given by
\begin{equation}
\langle t \rangle = \tau +t_f  {(1-p_s) \over p_s}
\label{eqn:SearchTime1}
\end{equation}
If we recall that $p_s=e^{-k_cd/v}$ and the result for $t_f$ in eqn.~\ref{eqn:tr}, we find
\begin{equation}
\langle t\rangle=\tau+\frac{1}{k_c} \frac{1-e^{-k_c \tau}(k_c \tau+1)}{1-e^{-k_c \tau}} \frac{\left(1 - e^{-k_c \tau}\right)}{e^{-k_c \tau}}.
\end{equation}
Noting the cancellations of the terms $1-e^{-k_c \tau}$ in the numerator and
denominator, this simplifies to the lovely final result
\begin{equation}
\langle t\rangle=\frac{1}{k_c}\left(e^{k_c \tau}-1\right).
\label{eqn:1DSearch}
\end{equation}
There are a variety of interesting limits to this result that are useful to consider.  For example,
in the case where the time to suffer a catastrophe  is much larger than $d/v$, the microtubule
will hit the chromosome on first try and the average time is $d/v$. To see this limit in practice when $k_c\tau \ll 1$,  we Taylor expand the
exponential resulting in
\begin{equation}
\langle t\rangle \approx \frac{1}{k_c}\left(1+k_c \tau-1\right) \approx \tau.
\end{equation}
Figure~\ref{fig:ChromosomeSearchTime1D} shows the average search
time as a function of the catastrophe rate $k_c$.  The minimum in this curve is
found when the catastrophe rate goes to zero. We can see this more formally by evaluating
\begin{equation}
\frac{d\langle t\rangle}{d k_c}=-\frac{1}{k_c^2}\left(e^{k_c \tau}-1\right)+\frac{\tau}{k_c} e^{k_c \tau}=0
\end{equation}
implying that the minimum occurs at
\begin{equation}
k_c = {1 \over \tau}\left(1-e^{-k_c\tau}\right),
\end{equation}
which is satisfied when $k_c\tau=0$,
meaning that the catastrophe rate itself is zero.
The best strategy for this particularly simple model is to have no catastrophes.  Let's now consider the more serious model that allows for a three-dimensional search~\cite{Holy1994,Wollman2005}.

\subsection{Three-Dimensional Chromosome Capture by Exploratory Dynamics}

The calculation given above made the unrealistic assumption that every microtubule started out
in the right direction, and thus we only needed to find out whether such microtubules lasted
long enough to reach the chromosome before suffering a catastrophe.  We now examine the case in which 
every time a microtubule starts anew, it has a probability $f_r$ of going in the right direction
to meet the chromosome, and a probability $f_w=1-f_r$ of going in the wrong direction. 
This concept is illustrated schematically in Figure~\ref{fig:ChromosomeSchematic}(C).  

\begin{figure}
\centering{\includegraphics[width=5.3truein]{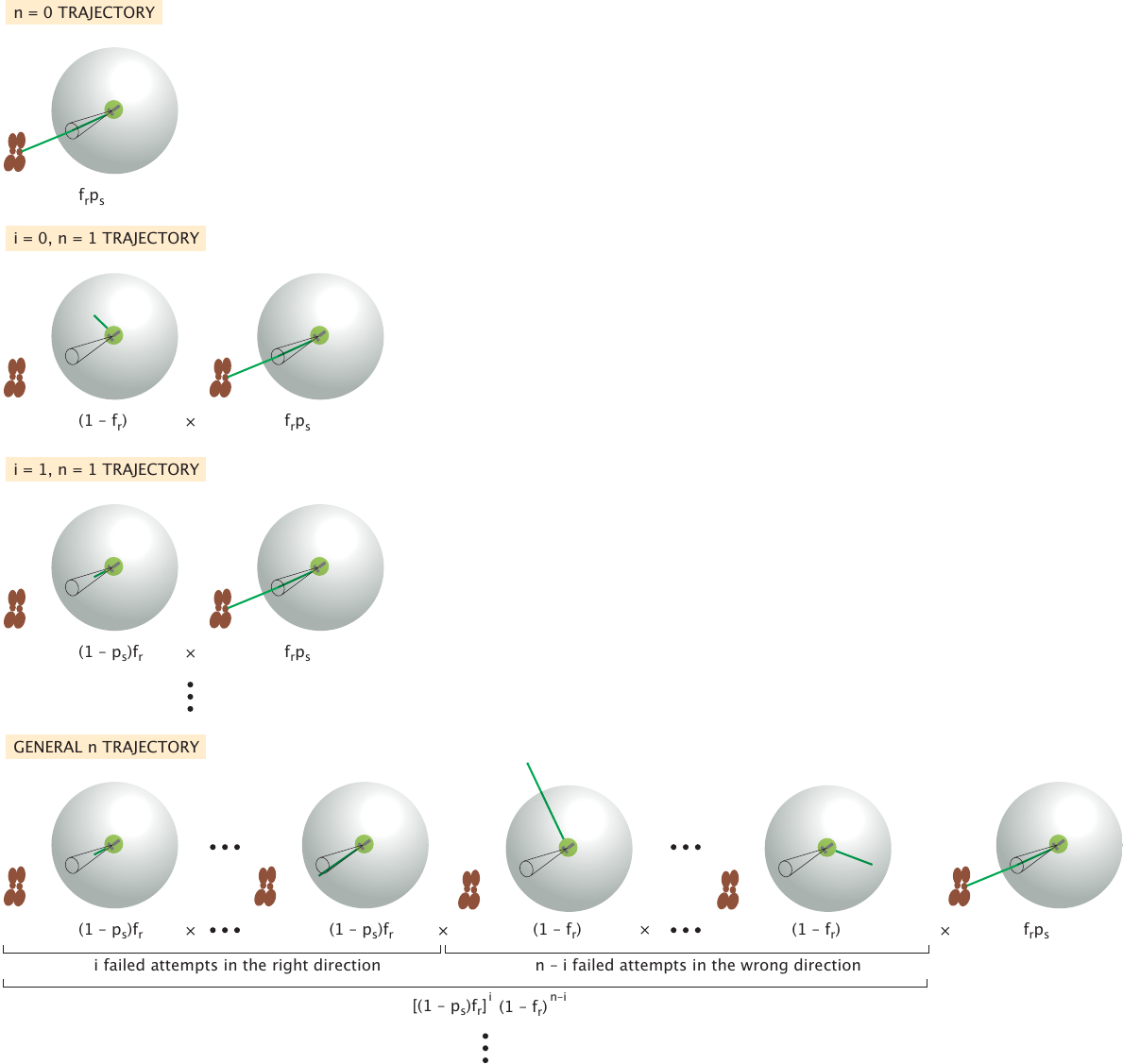}}
\caption{Trajectories and weights for three-dimensional model of chromosome search and capture.  Trajectories are labeled by the label $n$ which tells how many failed trajectories there were and the label $i$ which tells us out of the $n$ failed trajectories, how many of them were in the right direction. Each such trajectory has an associated probability and time until success and to find the average time we sum over the infinite set of trajectories as $n\rightarrow \infty$.
\label{fig:ChromosomeSearch3D}}
\end{figure}

As with the previous examples, we need to write an expression for the probabilities
of all the different kinds of trajectories as indicated in Figure~\ref{fig:ChromosomeSearch3D}.  For concreteness, let's consider those trajectories
that involve $n$ unsuccessful polymerization events before a success on the $(n+1)^{th}$ try.
However, there is a new twist relative to the calculation of the previous section.  Now our trajectories can fail
either because they are going in the right direction, but don't last long enough to reach the chromosome or
alternatively, they are going in the wrong direction and no matter how long they last, they will not
find the chromosome.  As a result, we introduce the notation $p_i(n)$ for those trajectories that fail $n$ times
before a success, with $i$ of those failures being in the right direction, but of insufficient duration,
and $n-i$ of those failures going in the wrong direction.   The probability of
a trajectory in the right direction, but that does not last long enough to reach the chromosome is
given by $f_r\left(1-p_s\right)$ while the probability of a trajectory in the wrong direction is given by
$\left(1-f_r\right)$. Given these partial probabilities, we can now write the probability of
 $p_i(n)$ of
this category of trajectories as
\begin{equation}
p_i(n)=\frac{n!}{i!(n-i)!} \underbrace{f_r^i\left(1-p_s\right)^i}_{{\mbox{right direction}}}\underbrace{\left(1-f_r\right)^{n-i}}_{{\mbox{wrong direction}}}\ f_r p_s,
\end{equation}
where we see that out of the $n$ unsuccessful trajectories, $i$ of them went
in the right direction and $n-i$ of them went in the wrong direction.
The prefactor counts the number of ways that out of these $n$ trajectories, $i$ of
them will be in the right direction and $n-i$ of them will be in the wrong direction.

The time associated with a trajectory involving $n$ unsuccessful polymerization events  before a final
successful polymerization event  is given by
\begin{equation}
t_i(n)=i \, t_r+(n-i) t_w+\tau.
\label{eq:timeToAvg_3dChromoCapture}
\end{equation}
Here we acknowledge that $i$ of the unsuccessful trajectories go in the right direction and have an average lifetime $t_r$, $n-i$ of the unsuccessful trajectories go in the wrong direction and have an average
lifetime $t_w$ and the final successful trajectory has a time $\tau$.
As a reminder, recall that 
\begin{equation}
t_w={1 \over k_c},
\end{equation}
since these trajectories are of unrestricted length and
$t_r$ is given by the expression in eqn.~\ref{eqn:tr} since these trajectories are constrained
to be less than $\tau=d/v$.

Now we need to sum over all possible trajectories which we can organize successively
by considering all trajectories with 0 failures, all with 1 failure, all with 2 failures $\cdots$ and all with $n$ failures and so on, resulting in 
\begin{equation}
\langle t\rangle=\sum_{n=0}^{\infty} \sum_{i=0}^n t_i(n)p_i(n).
\end{equation}
More explicitly, this can be written as
\begin{equation}
\langle t\rangle=\sum_{n=0}^{\infty} \sum_{i=0}^n \frac{n!}{i!(n-i)!} f_r^i\left(1-p_s\right)^i\left(1-f_r\right)^{n-i} f_r p_s 
\times\left[i t_r+(n-i) t_w+\tau\right].
\end{equation}

\begin{figure}
\centering{\includegraphics[width=5.3truein]{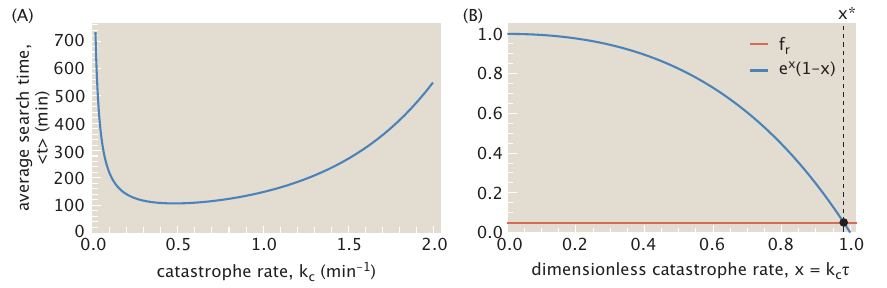}}
\caption{Optimal search time as a function of the catastrophe rate.  (A) Time to capture the chromosome as a function of the catastrophe rate. (B) Plot of the two sides of the equation for
the optimal search time (eqn.~\ref{eqn:OptimalSearchCondition}) as a function of the catastrophe rate $k_c$.   The parameters used here are: $f_r= 0.05$, $d=5~\mu $m, $v = 10~\mu$m/min implying $\tau = v/d = 2~$min.
\label{fig:ChromosomeSearchTime}}
\end{figure}

Using all of the tools already introduced throughout the paper, all of these sums are
of the form $\sum_{i=0}^{\infty} i x^i$ which we can evaluate to yield
\begin{equation}
\langle t\rangle=\tau+\frac{1}{p_s f_r}\left(t_r\left(1-p_s\right) f_r+t_w\left(1-f_r\right)\right)
\end{equation}
which simplifies to the remarkably compact result
\begin{equation}
\langle t\rangle=\frac{1}{k_c}\left(\frac{e^{k_c \tau}}{f_r}-1\right),
\label{eqn:3DsearchTime}
\end{equation}
nearly identical to the earlier result of eqn.~\ref{eqn:1DSearch}.
This expression is plotted in Figure~\ref{fig:ChromosomeSearchTime}(A)
where we see that there is an optimal catastrophe rate to achieve
finding the chromosomes in the least time.

It is of interest to explore this expression in various limits to see if it jibes with
our intuition.  We note that in the case when $f_r=1$, we recover
precisely the same result we found earlier as eqn.~\ref{eqn:1DSearch}, meaning
that in the case where every microtubule starts off in the right direction, the present
expression reduces to the earlier equation based upon the simplified model of
 Figure~\ref{fig:ChromosomeSchematic}(B). 
As we did before, we can examine this equation for the search time to see if we
 can understand where it takes its minimum with respect to the catastrophe rate.
 To that end, we evaluate
 \begin{equation}
\frac{d\langle t\rangle}{dk_c}=-\frac{1}{k_c^2}\left[\frac{e^{k_c \tau}}{f_r}-1\right]+\frac{1}{k_c}\left[\frac{\tau}{f_r} e^{k_c \tau}\right]=0
\end{equation}
which simplifies to the condition
\begin{equation}
\frac{1}{k_c^2}\left[\frac{e^{k_c \tau}}{f_r}-1\right]=\frac{\tau}{k_c f_r} e^{k_c \tau}.
\end{equation}
If we introduce the notation $x=k_c\tau$, then this condition simplifies further to
\begin{equation}
f_r=e^x(1-x).
\label{eqn:OptimalSearchCondition}
\end{equation}
In Figure~\ref{fig:ChromosomeSearchTime}(B), we plot both sides of this
equation and note that the solution corresponds to the point of intersection of
the two curves, the horizontal line representing the left side, $f_r = \mbox{constant}$ and the
curve corresponding to the right side $e^x(1-x)$.  However, since we know that $f_r \ll 1$ because the microtubules polymerizing in the ``right'' direction is rare, this licenses defining the small parameter 
$\varepsilon$ via  $x=1-\varepsilon$ resulting in
\begin{equation}
f_r=e^{(1-\varepsilon)}(1-(1-\varepsilon))
\end{equation}
Now because of the smallness of the parameter $\varepsilon$, we can Taylor expand
the exponential resulting in the very simple result
 \begin{equation}
\varepsilon \approx \frac{f_r}{e}.
\end{equation}
Recalling that $x=k_c\tau=1-\varepsilon$, we now have an approximate value for
the optimal catastrophe rate given by
 \begin{equation}
k_c \approx \frac{1}{\tau}\left(1-\frac{f_r}{e}\right).
\end{equation}
Given the fact that nearly all directions are ``wrong'' this implies that $f_r \ll 1$ (e.g. we used
$f_r=0.05$ in our plots), we end with the very simple result that the optimal catastrophe rate corresponds to $k_c \approx 1/\tau$.

Note that our treatment of the dynamics thus far has considered trajectories in which 
the lifetimes associated with the various processes were all average lifetimes.
A more careful treatment of the exploratory dynamics would sum over {\it all} possible
lifetimes for each and every step.  This would mean that instead of considering every
right or wrong trajectory as having a lifetime, $t_r$ or $t_w$, respectively, we would sum over {\it every}
allowed lifetime.  This would turn our sums over trajectories into integrals over trajectories, but ultimately
would not change the results obtained above.

\subsection{Molecular Control of Exploratory Dynamics}

The description of exploratory dynamics considered thus far in the paper is decidedly
incomplete. There is a whole pageant of accessory proteins that are part of
the process of chromosomal search and capture that we have not mentioned
at all since the primary focus of our paper is to focus on general ideas about exploratory dynamics. Several excellent reviews give a scholarly assessment of the complexities of
the process of chromosomal search~\cite{Heald2015, Lacroix2022}. Further,
several earlier theoretical approaches examined precisely these kinds of elaborations
on the most naive version of the search and capture model as exploratory dynamics~\cite{Holy1994,Wollman2005}. The way we think about the many molecular interventions implicated in the low error
rate construction of the spindle of dividing cells~\cite{Ha2024} is that the parameters $k_c$ and
$v$ used throughout our discussion are not fixed, but rather are subject to molecular control. For example,
we know that there are a variety of proteins that can either speed up or hinder
the polymerization of cytoskeletal filaments~\cite{Lacroix2022}, which the model accounts for by having the rate $v$ be a function of concentration and state of modification of these regulatory proteins.

\begin{figure}
\centering{\includegraphics[width=2.8truein]{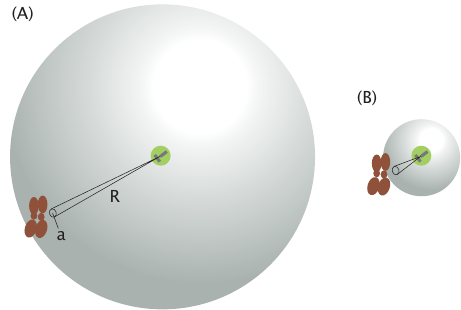}}
\caption{The geometry of chromosome search and cell size. (A) For a large cell size,
the cone of successful directions is characterized by the probability $f_r=a/4\pi R^2$. (B) For a smaller size, $f_r$ is larger.
\label{fig:ChromosomeSearchTimeCellSize}}
\end{figure}

This brief section will focus on how parameters such as the catastrophe rate $k_c$ and the microtubule growth rate $v$ might be controlled by specific molecular interventions. One reason for our interest in this topic is because we wonder how the search time depends upon the size of the cell itself, a topic that was examined carefully by Wollman et al.~\cite{Wollman2005}.
In particular, motivated by the idea that in some organisms during embryonic development
there are reductive cell divisions which continually reduce the size of the cells. This then has the effect of changing the distance between
chromosomes and the centrosome, which suggests that the search time will change as well, being shorter in smaller cells. Figure~\ref{fig:ChromosomeSearchTimeCellSize} illustrates
this idea by showing how the solid angle cone that characterizes growth in the ``right'' direction
depends upon cell size. As cells get smaller, the area of the correct solid angle becomes
a larger fraction of the total area.

To estimate how cell size, and hence distance between centrosome and kinetochore,
alters the search time, we begin by considering that the size of the kinetochore can be measured by an area $a$ (independent of the cell size) as shown in Figure~\ref{fig:ChromosomeSearchTimeCellSize}.  Hence, the fraction of trajectories which are starting in the right
direction is given by
\begin{equation}
f_r = {a \over 4 \pi R^2},
\end{equation}
where $R$ is a measure of the cell size or the distance from centrosome to kinetochore. %
In light of this parameter, we can rewrite the search time given in eqn.~\ref{eqn:3DsearchTime}
as
\begin{equation}
\langle t\rangle=\frac{1}{k_c}\left(\frac{4 \pi R^2}{a}e^{R / \lambda} -1\right),
\label{eqn:SearchTime3DTargetSize}
\end{equation}
where we have introduced the length scale
$\lambda=v/k_c$ which emerges naturally as a key parameter in the problem. Specifically, $\lambda$ enjoys the interpretation of the typical length grown by a microtubule before catastrophizing, so $R/\lambda$ is a dimensionless ratio characterizing how large the cell size is compared to typical growth excursions of the microtubule. An immediate and simple consequence of this model is that the search time is exponentially sensitive to changes in cell size, assuming that $R/ \lambda$ is large. In the opposite case, when $R/ \lambda$ is small, the exponential is roughly equal to one. In this case the search time grows quadratically with cell size. Either way, the simple model for kinetochore search we have introduced predicts that the search time will depend sensitively on the cell size.

Our examination of the simple model of kinetochore search is motivated primarily by the goal of revealing a larger set of problems that we feel are usefully described by the ideas of exploratory dynamics. In the specific context of chromosome capture the inadequacies of this model have been examined previously~\cite{Wollman2005}.   In particular, the model described above predicts search times that are significantly longer than those measured experimentally~\cite{Wollman2005}.  We now review the approaches that have been taken to account for
discrepancies with what is known experimentally. Here to demonstrate the utility of even this simple a model we start from the large body of literature,
very well reviewed in~\cite{Heald2015, Lacroix2022}, which explores the spatial
and temporal scaling of the spindles of dividing cells. As we will touch on below,
the simplest overarching observation is that the sensitive dependence of search time on cell size described
above is not consistent with the observed behavior.

Remarkably, in a variety of experiments on developing embryos and encapsulated {\it Xenopus} egg extracts, the search time is roughly constant (or more precisely, the spindle assembly time), independent of cell size~\cite{Lacroix2018,Rieckhoff2020}.  For a concrete
example, see Figure 6 of~\cite{Lacroix2018}.
In the embryo, different cell sizes result from reductive cell divisions, while in the synthetic system the experimentalist controls the size of the encapsulating vesicle, which plays the role of a synthetic cell.
These experimental observations on the lack of a cell-size dependence to the search time are at odds with the result of eqn.~\ref{eqn:SearchTime3DTargetSize}.  This fact invites us in the remainder of this section to consider how the two rate parameters, $v$ and $k_c$, should scale with the cell size~\cite{Belmont1990,Holy1994,Wollman2005,Lacroix2022} so that the search time will be independent of $R$. As cogently described by Belmont et al.~\cite{Belmont1990}, the dynamic instability model
features four rate parameters: the rates of polymerization and depolymerization, and the rates of
catastrophe and rescue. In the highly schematized model described here, we only preserved
two of those four rates, namely, the catastrophe rate $k_c$ and the polymerization rate associated with the speed of growth $v$,
since we ignored rescues altogether.

Based on eqn.~\ref{eqn:SearchTime3DTargetSize}, to prevent the search time from growing exponentially with cell size (when the cell is big compared to typical microtubule excursions so the dimensionless parameter $R/ \lambda$ is large), the typical length  the microtubules reach before collapsing must scale linearly with cell size. That is, $\lambda = v/k_c \propto R$, where the $\propto$ symbol represents the idea that if the quantity on the right hand side (i.e. $R$) were to be doubled or halved, the same fate would befall the quantity on the left hand side (i.e. $v/k_c$). We are still left with the term that is quadratic in $R$ and in our simple model we would further require $k_c \propto R^2$ in order for the search time to be independent of $R$. It is interesting to note that the only way to satisfy both scaling requirements is for the speed of polymerization to be proportional to $R^3$, which is the same as saying that it is proportional to the cell volume. This scaling is exactly what was found experimentally in~\cite{Lacroix2018}, and corresponds to a constant speed of polymerization per cell volume.

Another parameter that we have thus far not touched upon is the number of microtubules. Thus far, the entirety of our discussion has knowingly focused on the single microtubule case.
But of course, as is immediately obvious in the most cursory examination of dividing cells, there are many
microtubules present in the spindle~\cite{Baumgart2019}! More recently a set of beautiful experiments
revealed that the scaling of spindle size with cell size during embryonic development
is due to a finely-regulated number of microtubules in the spindle~\cite{Rieckhoff2020}. While our simple model does not consider this parameter,
the presence of multiple microtubules will decrease the search time in proportion to the number of microtubules, assuming that each one searches independently; this picture results in a revised search time for at least one microtubule to reach the kinetochore of,
\begin{equation}
\langle t\rangle=\frac{1}{Nk_c}\left(\frac{4 \pi R^2}{a}e^{R / \lambda} -1\right),
\label{eqn:SearchTime3DTargetSizeN}
\end{equation}
where $N$ is the number of microtubules.
Interestingly, this new twist allows us to return to the problem of how the search time can be independent of cell size. If the number of microtubules $N$ scales with the cell's surface area ($N \propto R^2$), the quadratic prefactor's dependence on $R$ drops out. Indeed, careful measurements have revealed precisely this kind of scaling~\cite{Rieckhoff2020}. The whole prefactor would  lose all cell size dependence if the catastrophe rate $k_c$ were constant with cell size, $k_c = \mbox{constant}$. Next, to further abolish the exponential factor's variation with cell size in eqn.~\ref{eqn:SearchTime3DTargetSizeN}, either the cell is small (giving $R/\lambda \ll 1$), or---as discussed previously---$\lambda=v/k_c \propto R$, which (with constant catastrophe rate) compels the microtubule growth rate $v$ to vary linearly with cell size $R$, namely $v \propto R$. These distinct biophysical relationships make precise predictions amenable to experimental scrutiny.
 
While we find these evocative comparisons between experimental data and the simple model of exploratory dynamics intriguing~\cite{Holy1994,Wollman2005}, much more
work needs to be done to specifically measure how the key parameters of models such as these scale with cell size (and other variables of physiological state). Here we only wish to point out how a simple model such as this points to fresh and specific questions about regulation of assembly of microtubules, which can be addressed by experiments.

\subsection{Exploratory Power}

\begin{figure}
\centering{\includegraphics[width=5.0truein]{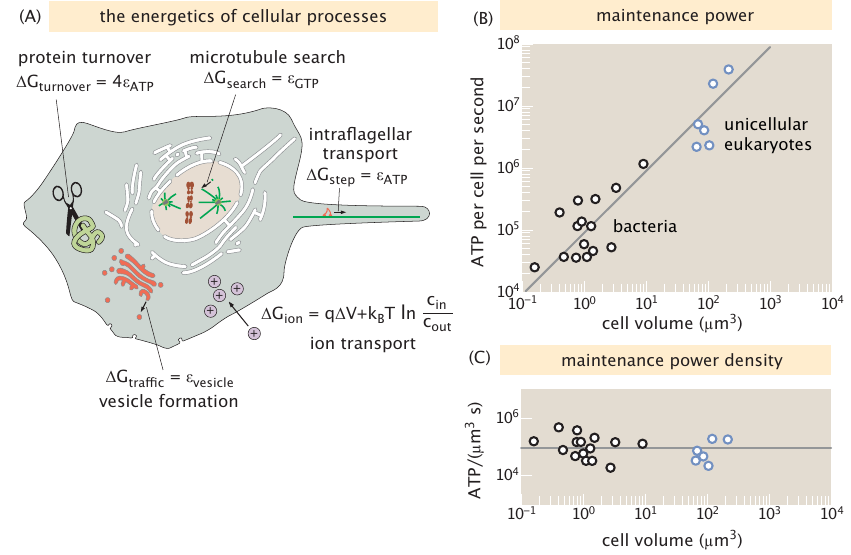}}
\caption{The power of biological processes. (A) Comparison of the bioenergetics of exploratory dynamics in chromosome search and capture with other key cellular processes.  In each case, the $\Delta G$ shows the free energy cost of a unit process such as the addition of a tubulin monomer to a growing microtubule or the cost of transporting a single ion across the cell membrane.  The power is given in turn by $J_{\text{process}}\Delta G_{\text{process}}$, where $J_{\text{process}}$ is the flux of the process of interest. (B) Power to maintain cellular processes, as inferred by continuous chemostat measurements, as a function of cell volume; proportional scaling (acceptably capturing these data) is denoted by the gray line~\cite{Marinov2015}. (C) Power densities per volume ensuing from the data in (B), reflecting a highly typical power density of $\langle \rho \rangle \approx 9 \times 10^5 ~\text{ATP}/(\mu \mbox{m}^3 \mbox{s})$ for overall cellular maintenance.}
\index{facilitated diffusion!theoretical model|ff}
\label{fig:KirschnerEnergyExploration}
\end{figure}

\begin{figure}
\centering{\includegraphics[width=3.0truein]{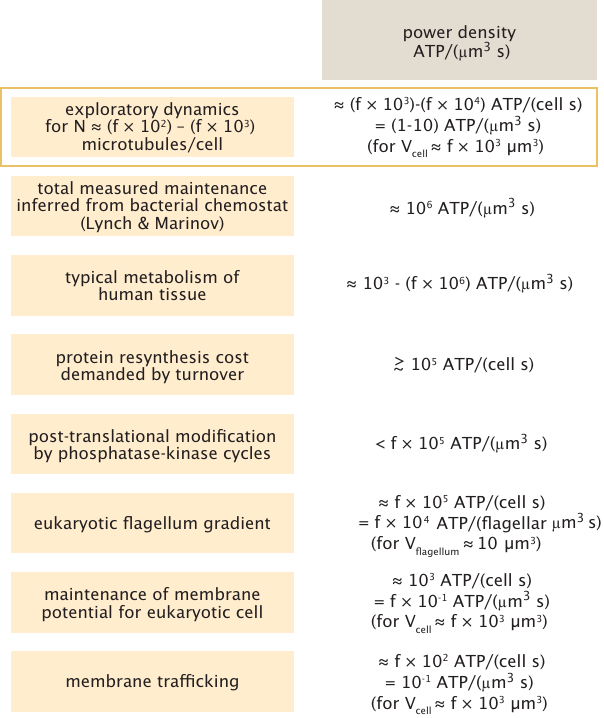}}
\caption{Summary of the power required for various cellular processes per volume. The goal of the figure is to compare the cost of microtubule search and capture by exploratory dynamics with the cost of other cellular processes.  Note that we use the symbol $f$ for ``few'' with the arithmetic rule that $f \times f \approx 10$~\cite{Mahajan2010}.}
\label{fig:tabulatedExploratoryDynamicsOoMestimates_NO2}
\end{figure}

At first glance, the analogy shown in Figure~\ref{fig:NYBostonRail} in which a railroad between Boston and New York is constructed by randomly laying down track seems
completely ridiculous.  When viewed at human engineering scales, exploratory dynamics
leads immediately to questions of efficiency, costs and benefits, words that are likely
too imprecise for discussing the molecular and cellular scale events that use this mechanism in living organisms.  
A fascinating question that arises in the context of these problems in exploratory dynamics is
how much they cost energetically. Post-translational modifications
imply a steady-state energy flux~\cite{Qian2007}; adaptation of the chemotaxis receptors a steady-state energy
flux in the form of methylation~\cite{Lan2012}; and the repeated cycles of growth and shrinkage in microtubule systems imply costs both in the form of cytoskeletal polymerization and of the motors that attend this process \cite{hill1982bioenergetics}.  Recent calorimetry measurements using microtubule-motor systems found a remarkably large heat dissipation in comparison with the larger scale mechanical motions of these collectives~\cite{Foster2023}.  In the remainder of this section, we perform a succession of calculations as shown in Figure~\ref{fig:KirschnerEnergyExploration} and summarized in Figure~\ref{fig:tabulatedExploratoryDynamicsOoMestimates_NO2},
beginning with the energetics of the exploratory dynamics of a single microtubule, followed
by crude estimates of the cost of all of the microtubules in a spindle and then a series of
comparisons of these powers and energies to other key processes in the dynamics of a cell.

\FloatBarrier

For the exploratory dynamics of microtubule search and capture
described in previous sections, we can make an elementary estimate of how many
ATPs or GTPs have been consumed. The simplest estimate asks us to examine the energetic cost
of repeated cycles of microtubule polymerization and depolymerization, ignoring for now the attendant costs associated with the many molecular motors that are part of the process as well.  The starting point of the estimate is the recognition that every tubulin dimer added to the growing microtubule costs a GTP as shown in  
Figure~\ref{fig:KirschnerEnergyExploration}(A) and is labeled as $\Delta G_{\text{search}}$.  
The total power is given as $J_{\text{search}}\Delta G_{\text{search}}$, where $J_{\text{search}}$ is the number of
tubulin additions per unit time.  As a result,
the rate of GTP hydrolysis can be estimated as 
\begin{equation}
J_{\text{search}}= {n_{\text{proto}} v \over l_{\text{mono}}}
\end{equation}
where $n_{\text{proto}}$ is the number of protofilaments in a microtubule, $v$ is the speed of microtubule growth in units of length/time and
$l_{\text{mono}}$ is the length of a tubulin dimer.  If we assign a characteristic energy of $\varepsilon_{GTP}=$20~$k_B$T per GTP hydrolysis, this tells us that the power is given by
\begin{equation}
\mbox{power} \approx J_{\text{search}}\varepsilon_{GTP}  \approx  \varepsilon_{GTP}{n_{\text{proto}} \, v \over l_{\text{mono}}}.
\end{equation}
If we approximate the relevant numbers such as $n_{\text{proto}} \approx 10$, $v\approx 10~\mu$m/min and $l_{\text{mono}} \approx~5~nm$, we find that the power for a single microtubule to perform exploratory behavior  is
roughly 500~ATP/s.

The results given above examine the energetics of a single microtubule doing repeated
instances of exploratory search and capture. But as any microscopy image will reveal,
this process involves hundreds to thousands of microtubules~\cite{Wollman2005,Kiewisz2022} and hence our estimates
both for the time scale of search and capture, and for the energetics and power need to
be amended to account for this effect.  As a result, we estimate the power expended
during the search and capture process is between $5 \times 10^3$ and $5 \times 10^4$~ATP/s.

With this equation in hand, we can also then estimate the total energy expended in the
search process given by the power times the elapsed time.
If we now recall eqn.~\ref{eqn:3DsearchTime} which tells us on average how long the search 
process lasts, we find that the total energy is given by,
\begin{equation}
\mbox{energy}=\text{power}\times \langle t \rangle = \varepsilon_{GTP} \times J_{\text{search}}\times \langle t \rangle = \frac{n_{\text {proto }} v\, \varepsilon_{GTP}}{l_{\text {mono }}} \frac{1}{k_c}\left[\frac{e^{k_c \tau}}{f_r}-1\right].
\end{equation}

Given these various estimates, it is of great interest to compare them to  other estimates of the energy budgets of cells.  A critical question that we come back to over and over in the context of
biological problems is how to judge whether a given process is ``costly'' or not.   The starting
point for answering that question is to take stock of the entire energy budget of a cell.  A systematic study was made of
the ``biosynthetic cost of a gene'' which attempted to work out the energetics of all of
the key processes that take place within cells of all types~\cite{Marinov2015}. In their
Figure 1 whose data we reproduce here as Figure~\ref{fig:KirschnerEnergyExploration}(B), they report on the maintenance cost with a value of roughly $f \times 10^{9}$~ATP/($\mu \mbox{m}^3$~hr), which translates into roughly $10^6$~ATP/($\mu \mbox{m}^3$~s). This characteristic power density, inferred from extrapolated chemostat measurements on bacteria and amoebozoa, also aligns with typical metabolic rates of various human tissue cells, ranging from the slower metabolism of resting erythrocytes ($\approx 8.5\times 10^{-17} \text{W}/\mu \text{m}^3$) to the rapid metabolism of T-cell lymphocytes in antigen responses ($\approx 6.5\times 10^{-13} \text{W}/\mu \text{m}^3$) \cite{freitas1999nanomedicine}, where we invoked the characteristic scale that one ATP hydrolysis liberates $10^{-19}$ J. Accordingly, a power density of about a million ATPs consumed per second per cubic micron serves as a very useful baseline power for the entirety of the processes associated with cellular maintenance.
To develop specific intuition, we consider several particular processes that are constantly in play in cells of different types and that will give us a sense of the scale of the cost of the search and capture process.

Cells constantly remake proteins to compensate for their active or spontaneous degradation. The energetic cost to resynthesize proteins can be formidable, as a simple estimate reveals. Specifically---assuming that the cellular abundance of proteins is maintained at a steady-state---the stable copy number $N$, overall resynthesis rate $q$, and degradation rate $k$, and typical lifetime $\tau$ can be related by Little's theorem in queuing theory \cite{little1961proof}, $N/\tau = q = k$.
Since mammalian cells typically have concentrations $c \approx$ a few million proteins per cubic micron \cite{milo2013total}; typical protein turnover rates in bacteria \cite{jayapal2010multitagging} and eukaryotes \cite{eden2011proteome} are both of order $k\approx f \times 10^{-5}~ \mbox{s}^{-1}$ (consistent with measured typical half lives of a few hours and where we remind the reader that we adopt the convention that $f = \mbox{few}$~\cite{Mahajan2010}); each protein is $\ell \approx $ a few hundred amino acids long and each peptide bond requires $\eta \approx$ 4 ATP equivalents to polymerize \cite{stephanopoulos1998metabolic}, protein resynthesis to maintain the standing proteome pool should demand of order $c k \ell \eta \approx 10^5$ ATP/($\mu \mbox{m}^3$~s). This significant cost can be within a factor few of the typical total power maintenance density of the cell we referred to above and is clearly much larger than the cost of the microtubule search and capture dynamics.  Proteins have energetic demands in other ways as well.  The post-translational modification of proteins can also demand modest to appreciable energy expenditures in a cell. Classic measurements and modeling of cyclic (de)phosphorylation cascades \cite{shacter1984energy} report that typical operating conditions for these cascades demand at most of order 1 mM ATP/min, corresponding to at most a $\text{few}\times 10^5$~ATP/$\mu$m$^3$~s.

Another fascinating, crucial, energetically-costly process that must be carried out at all times
is the maintenance of the membrane potential as seen in Figure~\ref{fig:KirschnerEnergyExploration}(A).  In particular, the difference in
ionic concentrations across membranes will be lost in the absence of active processes
to maintain those gradients.   Specifically, despite a very small permeability to such ions,
there is a leak current of ions across lipid bilayer membranes.   In particular, if we assume a typical permeability of
$p=10^{-5}$~nm/s~\cite{Milo2016} and a concentration difference across the membrane of
100~mM, then the flux is given by 
\begin{equation}
J_{ion}=p \Delta c \approx 1~{\mbox{ion} \over \mu \mbox{m}^2~\mbox{s}}.
\end{equation}
For a eukaryotic cell with a surface area of $f \times 10^3~\mu \mbox{m}^2$, this corresponds 
to $f \times 10^3$ ions/s.  The cost to transport one such ion up its gradient
is roughly $\Delta G_{ion} = q \Delta V + k_BT~\mbox{ln} (c_{in}/c_{out}) \approx 1$~ATP as seen in Figure~\ref{fig:KirschnerEnergyExploration}(A), indicating that the power of this process is about 10$^3$ ATP/s, a tiny fraction of
the overall power of the cell, but quite comparable to the power of a single microtubule engaging in
the search and capture process.

Membrane potentials are not the only gradients within cells.  Protein gradients of various kinds are
present in cells and embryos of all types.  A particularly well studied example that might give us
intuition for power and energy scales of such gradient formation is offered by
the eukaryotic flagellum. Here the process of intraflagellar transport (IFT) is characterized by molecular motors that carry cargo to the tip of the flagellum~\cite{Ishikawa2011} as seen in Figure~\ref{fig:KirschnerEnergyExploration}(A) with the cost of a single motor step given by $\Delta G_{step}=\varepsilon_{ATP}$.
This process results in a linear concentration gradient of molecular motors within the flagellum, some large fraction of which 
are attached to microtubules and thereby  consuming ATP~\cite{Chien2017}.  Rough estimates yield a power of the
form 
\begin{equation}
\mbox{power} = {c_0 +c_{\text{tip}} \over 2} L \left( {v \over \delta}\right) \varepsilon_{ATP},
\end{equation}
where $L$ is the length of the flagellum separating base from tip; $c_0$ and $c_{\text{tip}}$ are respectively the concentrations of motors at the base and tip; and $\delta$ is the step size of the motor.  Using  parameters reported by the experiments of ref.~\cite{Chien2017} yields $5 \times 10^5$~ATP/s, or $5 \times 10^4$~ATP/$\mu$m$^3$~s (on the basis of flagellar volume).

The process of membrane trafficking is constantly rearranging the internal and external membranes of cells in a highly energetically costly way.  As seen in Figure~\ref{fig:KirschnerEnergyExploration}(A), every time a vesicle is created, a simple estimate implies a cost of $\Delta G_{\text{traffic}} \approx$~10-20~ATPs corresponding to the 250-500~$k_BT$ cost of each such vesicle~\cite{Phillips2013}.  As was noted in a review on vesicles and trafficking:  ``A fibroblast kept in resting conditions in a tissue culture plate internalizes an amount of membrane equivalent to the whole surface area of the cell in one hour''~\cite{Kirchhausen2001}.  If we estimate the area of such a cell as $f \times 10^3~\mu \mbox{m}^2$ and consider 100~nm vesicles,  the area of each such vesicle is roughly $f \times 10^4~\mbox{nm}^2$, meaning roughly $6 \times 10^4$ vesicles are created every hour, or roughly 20 every second.  At a cost of 10-20~ATP per vesicle, this amounts to a power of roughly $f \times 10^2$~ATP/s.

All of these estimates---collectively summarized in Figure~\ref{fig:tabulatedExploratoryDynamicsOoMestimates_NO2}---form a useful and fascinating backdrop for trying to better understand
the energy budgets of cells and specifically, the costs of exploratory dynamics.  For a process
such as microtubule search and capture which results in the high-fidelity segregation of
chromosomes, even in light of such estimates, it is not clear how to decide if
a given power (or energy) is costly since chromosome segregation with errors can
be deadly, the highest cost of all.

\FloatBarrier

\section{Transcription Factor Search for Binding Sites as Exploratory Dynamics}
\label{section:TFSearch}

\begin{figure}
\centering{\includegraphics[width=3.0truein]{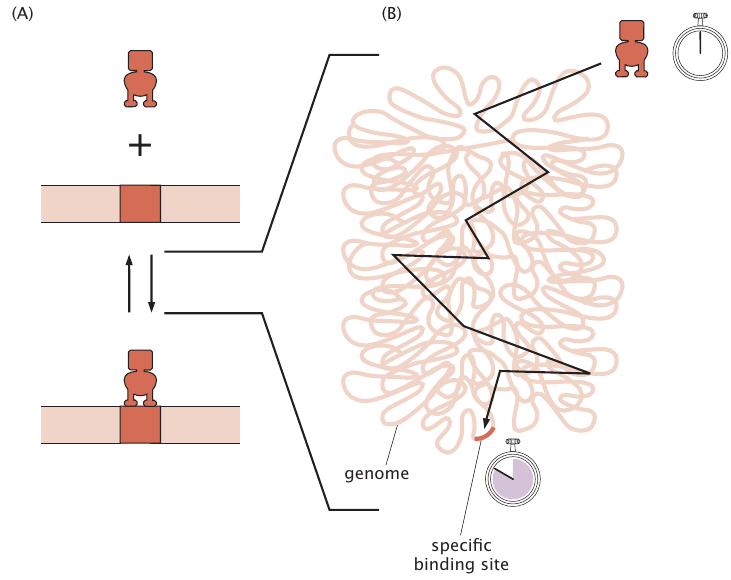}}
\caption{Searching for a binding site on the chromosome. (A) A particular binding site for a transcription factor  controls the expression of some gene of interest and our goal is to replace the simple schematized reaction on the left with a picture of the dynamics that acknowledges how the kinetics depends upon the complex geometry of the chromosome. (B) The transcription factor has to ``search'' throughout the entire genome. }
\index{facilitated diffusion!theoretical model|ff}
\label{fig:TFSearchDynamicsIntroduction2}
\end{figure}

\begin{figure}
\centering{\includegraphics[width=3.0truein]{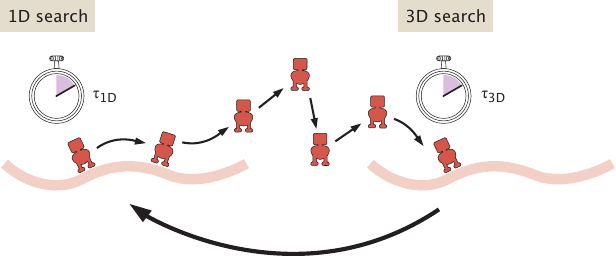}}
\caption{Simple model of combined 1D and 3D transcription factor search for a specific binding site. A transcription factor diffuses along the DNA for an approximate time $\tau_{1\mbox{D}}$ after which it falls of the DNA and performs three-dimensional diffusion for a time scale $\tau_{3\mbox{D}}$. The cycle is then repeated until the transcription factor reaches its binding site.}
\index{facilitated diffusion!theoretical model|ff}
\label{fig:1D3DSearchModelCartoon}
\end{figure}

The problem of how transcription factors search and find their binding sites provides another illuminating example of
exploratory dynamics.  When we write simple chemical reaction schemes for describing the binding and unbinding of transcription factors to their cognate sites as is shown in Figure~\ref{fig:TFSearchDynamicsIntroduction2}(A) they mask a much more complicated
underlying exploratory reality.  As seen in Figure~\ref{fig:TFSearchDynamicsIntroduction2}(B),   for a transcription factor to find its binding site and bind to it, the protein needs to traverse the complex intracellular environment including the complex polymeric geometry of the
chromosomal DNA itself.  Mechanistically, this exploratory dynamics takes place as shown
schematically in Figure~\ref{fig:1D3DSearchModelCartoon}, where a combination of
1D diffusion along the DNA punctuated by episodes of 3D diffusion result in a search time
that, in principle, is faster than either approach by itself.  We also note as was said above, that this mechanism shares many features of chromosome search and capture, but unlike that case, does not involve any energy expenditure.

Our approach to this problem essentially imitates a beautiful
treatment of the problem by Hachmo and Amir~\cite{Hachmo2023} though it is important to note that a broad swath of theoretical works have shed light on this fascinating topic of
``facilitated diffusion''~ \cite{vonHippel1979,Winter1981a,Winter1981b,Hu2006}. We invoke a model where a transcription factor can ``slide'' along the DNA and also jump between genomically distant but spatially proximal parts of the genome.
We note that, in addition to now classic in vitro bulk studies \cite{Winter1981a,Winter1981b}, there have been a variety of beautiful in vitro studies using the tools of single-molecule biophysics to watch proteins as they bind to and translocate along DNA~\cite{Wang2006,Blainey2006,vandenBroek2006,Finkelstein2010}.  Similarly, a new generation of single-molecule experiments has made it possible to query this same search process in the much more complex in vivo environment of living cells \cite{Hammar2012,Normanno2015}.  For our purposes, all of this work provides a rich canvas for further elaborating the general idea of exploratory dynamics.

\begin{figure}
\centering{\includegraphics[width=5.0truein]{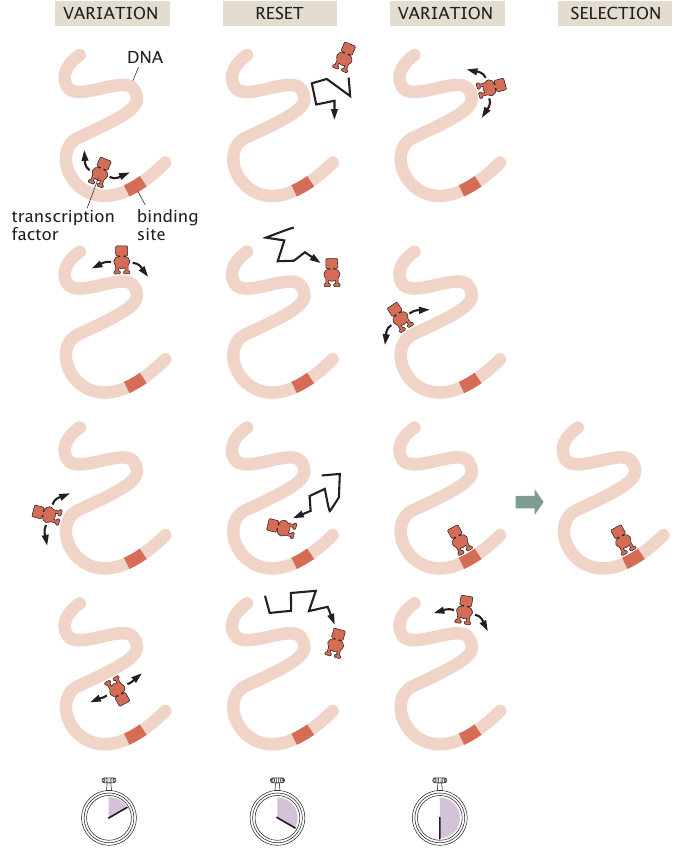}}
\caption{Exploratory dynamics of transcription factor search as an example of variation and selection. We can think of transcription factor search as a succession of search events along the DNA, punctuated by reset events of three dimensional diffusion.  ``Selection'' occurs when the transcription factor binds to its target site.}
\label{fig:KirschnerTFSearchClocks}
\end{figure}

In Figure~\ref{fig:1D3DSearchModelCartoon} we examine this canonical model of ``facilitated diffusion'' that combines 3D diffusion and 1D sliding along the DNA.   In this model, a transcription factor diffuses along the DNA with a diffusion constant $D_{1\mbox{D}}$ until it falls off the DNA after a time $\tau_{1\mbox{D}}$.  Once in the cytoplasm, the transcription factor undergoes 3D diffusion for a time $\tau_{3\mbox{D}}$ until it lands again on a different part of  the DNA.  Our goal is to estimate the search
time under this composite exploratory dynamics which can be thought of as one-dimensional localized
search for the binding site coupled with a ``reset'' mediated by three-dimensional diffusion and indicated
schematically in Figure~\ref{fig:KirschnerTFSearchClocks} using the evocative language of
variation (diffusion) and selection (binding).  

We begin by examining the probability of the transcription factor finding its target solely through 1D diffusion. The key idea in formulating the estimate is that during the time the transcription factor engages in 1D diffusion, it covers a distance given approximately by
\begin{equation}
\label{eq:1DSlidingLength}
	L_{1\mbox{D}} = \sqrt{D_{1\mbox{D}} \tau_{1\mbox{D}}}.
\end{equation}
If the total length of DNA to be searched is $L_{\mbox{genome}}$ and the transcription factor lands on a random DNA position, then the probability $p$  that the transcription factor will find its specific binding site during an episode of 1D diffusion that lasts a time
$\tau_{1\mbox{D}}$ is given by the fraction of the genome explored during a sliding period.   This argument leads to the estimate
\begin{equation}\label{eq:1DNoHopProb}
	p = {L_{1\mbox{D}} \over L_{\mbox{genome}}} = \sqrt{D_{1\mbox{D}} \tau_{1\mbox{D}} \over L_{\mbox{genome}}^2}.
\end{equation}

Now we want to compute the probability of finding the site during repeated cycles of 1D and 3D diffusion.   First, as shown in eqn.~\ref{eq:1DNoHopProb}, the probability of finding the binding site after the first sliding period, corresponding to zero 3D hops, is given by
\begin{equation}\label{eq:TFSearchp}
	p_0 = p = \sqrt{D_{1\mbox{D}} \tau_{1\mbox{D}} \over L_{\mbox{genome}}^2}.
\end{equation}
This takes approximately a time
\begin{equation}
	T_0 = \tau_{1\mbox{D}}.
\end{equation}
If no site is found during the first sliding interval, there is a chance that the site will be found during the next sliding period. This probability is given by
\begin{equation}
	p_1 = (1-p) p,
\end{equation}
where $1-p$ is the probability of the first 1D search being unsuccessful, and $p$ the probability of succeeding in the next 1D search. The time it takes for a binding site to be found in this second attempt, after one 3D hop, is given by
\begin{equation}
	T_1 = \tau_{1\mbox{D}} + \tau_{3\mbox{D}} + \tau_{1\mbox{D}},
\end{equation}
where the first term corresponds to the time of the first 1D search, the second term to the time of the subsequent 3D search, and the last term to the time of the second 1D search, when the site was found. Similarly, if the binding site is found after the second 3D hop, the transcription factor will have to have failed in its search during the the previous two 1D searches such that 
\begin{equation}
	p_2 = (1-p) (1-p) p
\end{equation}
and
\begin{equation}
	T_2 =  \tau_{1\mbox{D}} + \tau_{3\mbox{D}} + \tau_{1\mbox{D}} + \tau_{3\mbox{D}} + \tau_{1\mbox{D}}.
\end{equation}
We see a clear pattern emerge, namely, for the case involving $i$ episodes of 3D search the probability of finding
the site is given by
\begin{equation}\label{eq:pi1D3DSearch}
	p_i = (1-p)^i p
\end{equation}
and the corresponding search time is
\begin{equation}
	T_i = (i+1) \tau_{1\mbox{D}} + i \tau_{3\mbox{D}}.
\end{equation}
As a result, we can rewrite the average time for the transcription factor to find its binding site as
\begin{equation}\label{eq:TFSearchAverageTime1}
	T = \sum_{i=0}^{\infty} T_i p_i,
\end{equation}
where $p_i$ is the probability of the transcription factor finding its site after $i$ 3D hops, and $T_i$ is the time it takes to perform the trajectory with $i$ episodes of 3D diffusion. 
We can write this all out explicitly as 
\begin{equation}\label{eq:TFSearchAverageTime2}
	T = \sum_{i=0}^{\infty} T_i p_i = \sum_{i=0}^{\infty} (1-p)^i p \, \left((i+1) \tau_{1\mbox{D}} + i \tau_{3\mbox{D}} \right).
\end{equation}
We note that the mathematical structure of this problem is essentially identical to our calculations
in the previous section where we also found that our exploratory trajectories have probabilities
described by the geometric distribution.  This means that we can exploit much of the
same algebra we did in the previous section.

To start, we expand the terms in parentheses in eqn.~\ref{eq:TFSearchAverageTime2} to obtain
\begin{equation}\label{eq:TFSearchAverageTime3}
	T = \tau_{1\mbox{D}} \, p    \sum_{i=0}^{\infty}(1-p)^i i+ \tau_{1\mbox{D}} \, p \sum_{i=0}^{\infty}(1-p)^i + \tau_{3\mbox{D}} \, p \sum_{i=0}^{\infty}(1-p)^i i.
\end{equation}
We now regroup the terms and arrive at
\begin{equation}\label{eq:TFSearchAverageTime4}
	T = (\tau_{1\mbox{D}} + \tau_{3\mbox{D}}) p \sum_{i=0}^{\infty}(1-p)^i i + \tau_{1\mbox{D}} \, p \sum_{i=0}^{\infty}(1-p)^i.
\end{equation}
We note that the second sum is of the form of a geometric series which can be summed once again to yield
\begin{equation}
	\sum_{i=0}^{\infty}(1-p)^i = {1 \over 1 - (1-p)} = {1\over p}
\end{equation}
As a result, we have
\begin{equation}
	T = (\tau_{1\mbox{D}} + \tau_{3\mbox{D}}) p \sum_{i=0}^{\infty}(1-p)^i i + \tau_{1\mbox{D}} \, p {1\over p} = 
	(\tau_{1\mbox{D}} + \tau_{3\mbox{D}}) p \sum_{i=0}^{\infty}(1-p)^i i + \tau_{1\mbox{D}}.
\end{equation}
To carry out the remaining sum, we invoke the use of the derivative trick we originally introduced in eqn.~\ref{eqn:DerivativeTrick}.   Namely, we write the sum as
\begin{equation}
	\sum_{i=0}^{\infty}(1-p)^i i = -(1-p)  \frac{d}{d p} \left( \sum_{i=0}^{\infty}(1-p)^i\right),
\end{equation}
where the extra $-(1-p)$ term takes care of the fact that, after taking the derivative, the $(1-p)^i$ term inside the summation would otherwise become $-(1-p)^{i-1}$. We again have the familiar geometric sum that can be carried out to yield
\begin{equation}
	-(1-p)  \frac{d}{d p} \left( \sum_{i=0}^{\infty}(1-p)^i\right) = 
	-(1-p) \frac{d}{d p} \frac{1}{p}=-(1-p) \frac{(-1)}{p^2} = \frac{1-p}{p^2}.
\end{equation}
As a result, eqn.~\ref{eq:TFSearchAverageTime4} becomes
\begin{equation}
	T =  (\tau_{1\mbox{D}} + \tau_{3\mbox{D}}) p {1-p \over p^2} + \tau_{1\mbox{D}} \, p {1\over p} =
	(\tau_{1\mbox{D}} + \tau_{3\mbox{D}}) \frac{1-p}{p} + \tau_{1\mbox{D}}.
\end{equation}
This can be simplified to the form
\begin{equation}
	T = {\tau_{1\mbox{D}} \over p} + {\tau_{3\mbox{D}} \over p} - \tau_{3\mbox{D}}.
\end{equation}
Now, we assume that $p\ll1$, meaning that it will always take many cycles of 1D and 3D exploration to find the binding site. Mathematically, this means $\tau_{3\mbox{D}}/p \gg \tau_{3\mbox{D}}$ such that we can ignore the last $\tau_{3\mbox{D}}$ term in the expression. In light of this approximation, we find a  total search time of the form
\begin{equation}\label{eq:TFSearchAverageTime5}
	T = {\tau_{1\mbox{D}} \over p} + {\tau_{3\mbox{D}} \over p}.
\end{equation}

To understand the implications of the calculation that led to the total search time shown in eqn.~\ref{eq:TFSearchAverageTime5}, we recall the definition of $p$ introduced in eqn.~\ref{eq:TFSearchp}. As a result of this definition, we can use the formula for $p$ to  rewrite our result for the search time as
\begin{equation}
	T =  \sqrt{\tau_{1\mbox{D}} L_{\mbox{genome}}^2 \over D_{1\mbox{D}} } + \tau_{3\mbox{D}} \sqrt{L_{\mbox{genome}}^2 \over D_{1\mbox{D}} \tau_{1\mbox{D}}}.
\end{equation}
Further, we will non-dimensionalize the equation by measuring the search time in units of the 3D exploration time $\tau_{3\mbox{D}}$, leading to
\begin{equation}
	{T \over  \tau_{3\mbox{D}}} = \bar{T} =  \sqrt{{\tau_{1\mbox{D}} \over \tau_{3\mbox{D}}}{ L_{\mbox{genome}}^2 \over D_{1\mbox{D}} \tau_{3\mbox{D}} }} +  \sqrt{L_{\mbox{genome}}^2 \over D_{1\mbox{D}} \tau_{1\mbox{D}} },
\end{equation}
where we have defined $\bar{T} = {T / \tau_{3\mbox{D}}}$. In addition, we define the non-dimensional 1D search time  $\bar{\tau}_{1\mbox{D}} = {\tau_{1\mbox{D}} / \tau_{3\mbox{D}}}$ measured in
units of the 3D search time,  to arrive at
\begin{equation}
	\bar{T} =  \sqrt{ \bar{\tau}_{1\mbox{D}} { L_{\mbox{genome}}^2 \over D_{1\mbox{D}} \tau_{3\mbox{D}} }} +  \sqrt{L_{\mbox{genome}}^2 \over D_{1\mbox{D}} \tau_{3\mbox{D}} \bar{\tau}_{1\mbox{D}}}.
\end{equation}
We see that both square roots in the equation have a factor
\begin{equation}\label{eq:alphaDefinition}
	\alpha =  \sqrt{{D_{1\mbox{D}} \tau_{3\mbox{D}} \over L_{\mbox{genome}}^2}} ,
\end{equation}
corresponding to the fraction of the genome the transcription factor would have explored by 1D diffusion during the time it was actually diffusing in 3D. As a result, the non-dimensionalized time can be written as
\begin{equation}
	\bar{T} = {1 \over \alpha} \left(  \sqrt{\bar{\tau}_{1\mbox{D}}}  + \sqrt{1 \over \bar{\tau}_{1\mbox{D}}} \right).
    \label{eq:searchTimeBalance}
\end{equation}

This expression for the search time makes the striking prediction that there is an optimal balance between the time the transcription factor spends searching in one dimension and three dimensions. Specifically, when the one-dimensional search time is relatively long ($\bar{\tau}_{1\mbox{D}}\gg 1$), the first term $\sqrt{\bar{\tau}_{1\mbox{D}}}$ dominates the overall search time, implying that  the overall search time grows as $\bar{T}\thicksim \bar{\tau}_{1\mbox{D}}^{1/2}$. Otherwise (for short one-dimensional search times $\bar{\tau}_{1\mbox{D}}\ll 1$), the second term $1/\bar{\tau}_{1\mbox{D}}\gg 1$ dominates, implying in turn that  the overall search time shrinks as $\bar{T}\thicksim \bar{\tau}_{1\mbox{D}}^{-1/2}$. The crossover between these behaviors gives a minimum overall search time, when these terms are comparable.

We can quickly foresee that the value of the one-dimensional search time $\bar{\tau}_{1\mbox{D}}$ that minimizes the overall search time is located at $\bar{\tau}_{1\mbox{D}}=1$ by remembering the arithmetic mean-geometric mean inequality, which states that for two nonnegative values $a,b$, $\frac{a+b}{2} \geq \sqrt{ab}$; taking $b=1/a$ gives $a+\frac{1}{a} \geq 2$, where the optimal value of equality requires $a=1/a$, or $a=1$. (Amusingly, the same optimization problem emerges when asking for the side lengths of a rectangle of unit area that has the smallest perimeter.)  Letting $a$ be $\sqrt{\bar{\tau}_{1\mbox{D}}}$ translates this result to mean that smallest overall search time occurs when $\sqrt{\bar{\tau}_{1\mbox{D}}}=1$. 
This optimum may also be seen by taking the derivative of $\bar{T}$ with respect to $\bar{\tau}_{1\mbox{D}}$
\begin{equation}
	{d \bar{T} \over d \bar{\tau}_{1\mbox{D}}} = {1 \over \alpha} {d \over d \bar{\tau}_{1\mbox{D}}} \left(  \sqrt{\bar{\tau}_{1\mbox{D}}}  + \sqrt{1 \over \bar{\tau}_{1\mbox{D}}} \right)= {1 \over \alpha} \left( {1\over 2} {1\over \bar{\tau}_{1\mbox{D}}^{1/2}} - {1\over 2} {1\over \bar{\tau}_{1\mbox{D}}^{3/2}}  \right) = 0,
\end{equation}
which is satisfied if $\bar{\tau}_{1\mbox{D}} = 1$, regardless of the numerical value of the parameter $\alpha$. This result carries the physical meaning that $\tau_{1\mbox{D}}=\tau_{3\mbox{D}}$: the shortest overall search time is reached when the times spent by the transcription factor exploring in one and three dimensions are equal. 
Equivalently, this optimum corresponds to the case where the ratio of the distance traveled in 1D ($L_{1D}$) and that in 3D ($L_{3D} \equiv \sqrt{D_{3D} \tau_{3D}}$) are matched to an appropriate balance of the respective diffusion coefficients, $\frac{L_{1D}}{L_{3D}} = \sqrt{\frac{D_{1D}}{D_{3D}}}$.

\begin{figure}
\centering{\includegraphics[width=4.5truein]{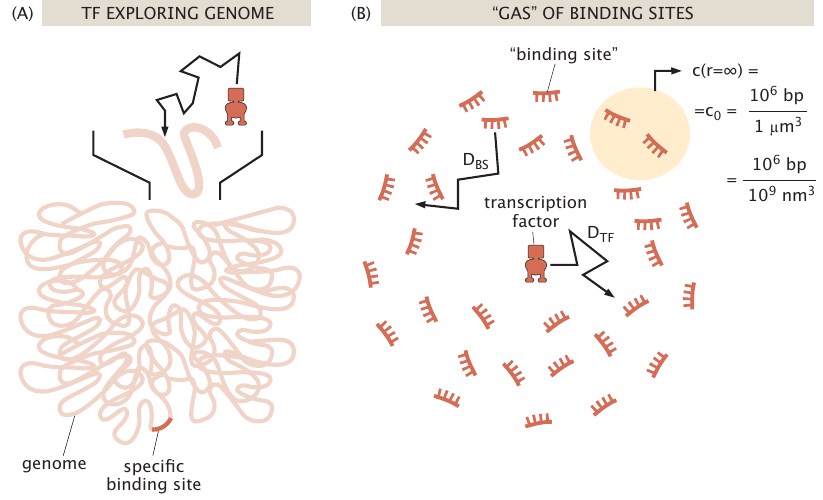}}
\caption{Estimating time scales of encounters between a transcription factor and the genome.  (A) During episodes of 3D diffusion, the transcription factor diffuses through the cellular environment before encountering the genome.  (B) Estimating the time scale of diffusion by using the diffusion-limited on rate idea by pretending that the binding sites themselves are diffusing around.  (C) Estimating the time scale of diffusion by pretending that the diffusion takes place within a discrete lattice of binding sites in 3D space.
\label{fig:tau3DEstimate}}
\end{figure}

Armed with this analytical understanding of the tradeoff between one dimensional searches and three dimensional resets, we now explore where typical cells operate parametrically. Before we can make progress on exploring all of these results numerically, we first need to estimate
some of the key parameters that are present in the model.  In particular, we need an estimate
of  $\tau_{3\mbox{D}}$, the time scale for 3D exploration between successive episodes of 1D diffusion along the DNA. Figure~\ref{fig:tau3DEstimate} offers a naive way to make such an estimate.  
In Figure~\ref{fig:tau3DEstimate}(B), we consider a transcription factor in a ``gas'' of binding sites. When the transcription factor is not bound to the DNA, we imagine these binding sites as freely diffusing with diffusion constant $D_{\mbox{bs}}$ in the vicinity of the transcription factor, and the transcription factor itself diffuses with a diffusion constant $D_{\mbox{TF}}$. In this simple model, the time $\tau_{3\mbox{D}}$ corresponds to the average time for one of these binding sites to reach the transcription factor. To estimate the concentration of this gas of binding sites, we use the fact that there are of order $10^6$ such binding sites in a bacterial genome in a volume of 1~$\mu$m$^3$, each one corresponding to a base pair in the genome.  We can use this concentration in conjunction with the diffusion limited on-rate \cite{Berg1993,Phillips2012} to estimate the diffusion time as 
\begin{equation}
\tau_{3\mbox{D}}={1 \over 4 \pi (D_{\mbox{bs}}+D_{\mbox{TF}})  a \, c_0}.
\end{equation} We note, however, that base pairs do not diffuse independently of each other. Because base pairs are part of the huge polymer that makes up the chromosome, their effective diffusion coefficient is expected to be much lower than that of a free base pair. Indeed, recent in vivo studies in a broad swath of organisms have reported an effective diffusion coefficient of chromosomal loci of the order of $10^{-2}~\mu m^2$/s \cite{Weber2010,Gabriele2022,Bruckner2023}. As a result, if we take the diffusion constant of a base pair as $D_{\mbox{bs}} \approx 10^{-2}~\mu\mbox{m}^2/s$ we see that it is much smaller than the diffusion constant of the protein itself ($D_{\mbox{TF}} \approx 10~\mu\mbox{m}^2/s$) and hence we can safely ignore the contribution of base pair diffusion.  If we now impose the  size of the  target site to be $a \approx f$~nm and the concentration of base pairs  is approximated by $c_0 = 10^6/\mu\mbox{m}^3 \approx 1 \mbox{mM}$, we arrive at $\tau_{3\mbox{D}}\approx f \times 10^{-6}$~s.
Finally, in order to estimate the numerical value of $\alpha$ defined in eqn.~\ref{eq:alphaDefinition}, we must also determine the 1D diffusion constant describing the sliding of the transcription factor along the DNA. Single-molecule studies have found 1D diffusion constants of the order of $D_{1\mbox{D}} \approx 10^5  - 10^6~\mbox{bp}^2/\mbox{s} \approx 10^{-2} - 10^{-1}~\mu m^2/s$ \cite{Wang2006,Blainey2006}. In light of all of these estimated parameters, we find $\alpha \approx f \times 10^{-7} - 10^{-6}$.

\begin{figure}
\centering{\includegraphics[width=3.0truein]{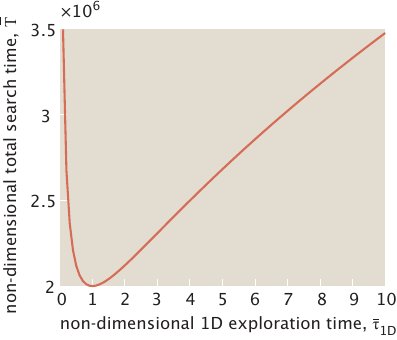}}
\caption{Time for a transcription factor to find its binding site in a model of combined 1D and 3D search.  The search time is measured in units of the time spent searching in 3D as a function of the relative time searching in 1D to 3D.  The parameter $\alpha$ was chosen to have the value $\alpha = 10^{-6}$ corresponding to the fraction of the genome that would have been explored by 1D diffusion during the time of a 3D exploration. Note that as the time spent exploring in 1D approaches zero, $\bar{\tau}_{1D}\rightarrow 0$, clearly the total overall search time $\bar{T}$ diverges, as encountering the correct binding site requires at least some time spent exploring in 1D.}
\label{fig:1D3DSearchOptimum}
\end{figure}

Using the parameters we have estimated, Figure~\ref{fig:1D3DSearchOptimum} shows the total search time as a function of the time spent searching in 1D in units of the time spent searching in 3D. The figure affirms the appearance and location of the optimal 1D search time, namely, when the time spent searching in 1D is equal to the time searching in 3D.

Despite how crude our estimates were, we can now examine whether the transcription factor binding site search dynamics takes a time consonant with the optimal search time which is found when the 1D and 3D search times are equal.   To make that determination, we need the 1D exploration time which we can then  compare to our estimate of the 3D exploration time that we performed earlier on in this section. In vivo experiments in bacteria have estimated the sliding length of a transcription factor to be of order $L_{1\mbox{D}} \approx 100$~bp \cite{Hammar2012}. Given our estimated 1D diffusion constant, this translates into a 1D search time of roughly $\tau_{1\mbox{D}}\approx 10^{-2}-10^{-1}$~s. This value for $\tau_{1\mbox{D}}$ 
differs by several orders of magnitude from our estimate of $\tau_{3\mbox{D}}$  shown in Figure~\ref{fig:tau3DEstimate} (B). However, it is important to note that neither the 1D nor the 3D exploration times have been directly measured in vivo and that our numerical estimates depend upon a succession of poorly known parameters. Indeed, recent experiments in mammalian cells paint a different picture. These experiments estimated that a transcription factor spends a similar amount of time in the 1D and 3D exploration modes (75\% of the time doing 3D exploration, to be more precise) \cite{Normanno2015}.

A second question we can explore guided by the estimate resulting in Figure~\ref{fig:1D3DSearchOptimum} is to ask whether this facilitated diffusion mechanism results in faster search times of transcription factors for their specific binding sites than explorations based on a 1D or 3D search alone. To make this possible, we first assign absolute units to the optimal non-dimensional total search time $\bar{T}$ shown in the figure. Specifically, we compute
\begin{equation}
    T_{\mbox{min}} = \bar{T}_{\mbox{min}} \tau_{3\mbox{D}} \approx f~\mbox{s}.
\end{equation}
Given in vitro measurements of the 1D diffusion constant of  $D_{1\mbox{D}} \approx 10^5  - 10^6~\mbox{bp}^2/\mbox{s}$ already described above, we can estimate a search time for a specific binding site through a genome composed of $L_{\mbox{genome}}=10^6$~bp as
\begin{equation}
T_{\mbox{1D}} \approx {L_{\mbox{genome}}^2 \over D_{1\mbox{D}}} \approx 10^6 - 10^7~\mbox{s},
\end{equation}
which is clearly several orders of magnitude slower than the search time associated with the  facilitated diffusion mechanism.
Finally, we estimate the  time for a search mechanism based on pure 3D diffusion as
\begin{equation}
T_{\mbox{3D}} = {1 \over 4\pi D_{\mbox{TF}}^{\text{eff}} a c_0},
\end{equation}
where $c_0 \approx 1$~nM, the concentration of an individual base pair in a $1\mu\mbox{m}^3$ nuclear volume, and where $a\approx f$~nm. To make this estimate, we also acknowledge that, in this case, $D_{\mbox{TF}}^{\text{eff}}$ is an effective diffusion constant that needs to account for the interactions between the transcription factor and the milieu of the nucleus such as those leading to crowding and to  non-specific DNA binding during the search process. Measurements and estimates of this effective diffusion constant remain an open subject. We choose values of $D_{\mbox{TF}}^{\text{eff}} = 10^{-1} - 1~\mu\mbox{m}^2/\mbox{s}$ based on recent in vivo measurements \cite{Normanno2015}. As a result, we get
\begin{equation}
T_{\mbox{3D}} = {1 \over 4\pi D_{\mbox{TF}}^{\text{eff}} a c_0} \approx f (10 - 10^2)~\mbox{s}.
\end{equation}
This search time is up to two orders of magnitude slower than the optimum search time due to facilitated diffusion $T_{\mbox{min}}$, suggesting that facilitated diffusion could be at play to speed up the transcription factor binding search process. Regardless of the specific numerical values, which are clearly still a subject of active investigation in the field, we posit that the model shown in this section is yet another intriguing example of the power of exploratory dynamics---in this case in the form of facilitated diffusion---to accomplish and potentially speed up specific biological functions.

\section{Discussion}
\label{section:Discussion}

The work of Prof. Erich Sackmann was characterized by a principled approach to understanding
the huge diversity of biological processes.  In the spirit of his many successes, we used
this opportunity to playfully engage with the idea that many biological processes
can be thought of as case studies in exploratory dynamics.  In contrast with the
conventional physics paradigms for dynamics in which, given some initial conditions,
the equations of dynamics are used to find the future evolution of the system, we have
argued that biology's unique and necessary ``solution'' to achieving some target
function is through a quite distinct alternative: exploratory dynamics.  From this point
of view, the classic ideas of  variation and selection in the context of biological evolution is hypothesized to have a much broader reach.

To provide some quantitative substance to the hypothesis of exploratory
dynamics, we considered several important case studies that have been studied
rigorously and successfully for decades, namely, chromosome search and capture, and
the search of transcription factors for their target DNA binding sites. 
Though our paper focused primarily on these particular case studies,  we argue that the idea of exploratory dynamics has much
broader biological reach, both in terms of the phenomena of living organisms and in the diversity of
possible theoretical approaches to these problems.

Ultimately, from a theoretical perspective, we think of questions in exploratory dynamics
as falling within the broad province of statistical physics. More specifically, there are a growing number of
examples of different approaches to considering the statistical mechanics of microscopic
trajectories as opposed to the more traditional statistical mechanics of states.  
The underlying statistical mechanics described here forced us to consider a number of problems involving
the geometric distribution.  However, we suspect that there are other mathematical paradigms
that will be useful, or even necessary, for considering the broad class of problems in exploratory
dynamics.  For example, there is a growing literature on the subject of random walks with reset in which
the random walker is forced to return to the origin~\cite{Evans2020}.

Clearly, molecular structures, cells, and organisms are often steered by external cues that vary in time, space, and history---cues that themselves are often sculpted by feedbacks from exploratory trajectories, as with ants leaving odor trails. Further, exploratory success probabilities (setting the ``life expectancies" of exploratory trajectories) may manifest rich memory effects, for instance when having explored for a while itself increases or decreases the probability of impending success. These complexities urge new mathematical frameworks to be developed and applied that accurately capture the exquisite and complex resulting tradeoffs and consequences of exploration. While it may prove productive to build metaphors between the biological phenomena showing these complexities and algorithmic phenomena such as reinforcement learning or stochastic gradient descent, our central contention is that the unique ingredients, constraints, and goals of biology demand altogether new classes of exploratory behavior whose structure and contrasting examples may also shed light on these more traditional domains.

In summary, we view a fascinating dichotomy between the many distinct and impressive dynamical
laws that have formed the backdrop of physics and mathematics for centuries and the unique
and necessary exploratory dynamics adopted by many biological processes.   Though we often adopt perspectives  in which systems move deterministically from some initial condition, 
from the examples of   exploratory dynamics, perhaps one learns that the best plan can be no plan at all.

\section*{Acknowledgments}

We are grateful to the NIH for support through award numbers DP1OD000217 (Director's Pioneer Award) and NIH MIRA 1R35 GM118043-01 to RP.  
MK acknowledges support from the NIH under grants R35-GM145248  and R01 AG07334.
JK was supported by the Simons Foundation and NSF grants DMR-1610737 and MRSEC DMR-2011846. 
HGG was supported by NIH R01 Awards R01GM139913 and R01GM152815, by the Koret-UC Berkeley-Tel Aviv University Initiative in Computational Biology and Bioinformatics, by a Winkler Scholar Faculty Award, and by the Chan Zuckerberg Initiative Grant CZIF2024-010479. HGG is also a Chan Zuckerberg Biohub Investigator (Biohub--San Francisco). GLS thanks the US NSF Graduate Research Fellowship Program under Grant DGE-1745301 and Caltech’s Center for Environmental Microbial Interactions for support. 
We are all deeply grateful to the CZI Theory Institute Without Walls for the support that makes our exploratory studies of exploratory dynamics possible.  We have benefited greatly from conversations with Xingbo Yang, Peter Foster, Tim Mitchison, Natalia Orlovsky, Lena Koslover, Yuhai Tu and Milo Lin.   Ana Duarte and Sara Mahdavi provided helpful discussion as well as numerical simulations.  We are also grateful to Ariel Amir for  very helpful discussions and insights and for suggesting the approach for transcription factor search.

\bibliography{PaperLibraryMathBiology}

\end{document}